\documentclass[a4paper,11pt]{article}
\pdfoutput=1 
\usepackage{jheppub} 

\usepackage[T1]{fontenc} 

\pdfoutput=1
\usepackage{amsmath,amssymb}
\usepackage{amsfonts}
\usepackage{bm,bbm}
\usepackage{graphicx}
\usepackage{mathrsfs}
\usepackage{slashed}
\usepackage{booktabs}
\usepackage{tabu}
\usepackage[dvipsnames]{xcolor}
\usepackage[normalem]{ulem}
\usepackage{tikz}
\usepackage{soul}

\usepackage{hyperref}
\hypersetup{colorlinks,citecolor= blue,linkcolor= blue, urlcolor=blue}

\usepackage{xcolor}
\usepackage{ulem}
\usepackage{array}
\usepackage{verbatim}
\usepackage{epsfig}
\usepackage{multirow}

\newcommand{\be}{\begin{equation}}
\newcommand{\ee}{\end{equation}}
\newcommand{\ba}{\begin{array}}
\newcommand{\ea}{\end{array}}
\newcommand{\bea}{\begin{eqnarray}}
\newcommand{\eea}{\end{eqnarray}}

\usepackage{ulem,fancyvrb}
\usepackage{xcolor}

\title{Low-Energy Supernova Constraints on Millicharged Particles}

\author[a]{Changqian Li,}
\author[a]{Zuowei Liu,}
\author[a]{Wenxi Lu,}
\author[a]{Zicheng Ye}

\affiliation[a]{Department of Physics, Nanjing University, Nanjing 210093, China}

\emailAdd{changqianli@smail.nju.edu.cn}
\emailAdd{zuoweiliu@nju.edu.cn}
\emailAdd{luwenxi@smail.nju.edu.cn}
\emailAdd{zichengye@smail.nju.edu.cn}

\abstract{

The hot and dense conditions of the supernova core provide an ideal environment for the production of new feebly-interacting particles. Low-energy supernovae, characterized by low explosion energy, are particularly intriguing due to their stringent constraints on  energy transfer from the core to the mantle by new particles. We investigate low-energy supernova constraints on millicharged particles by considering three production channels in the core: plasmon decay, proton bremsstrahlung, and electron-positron annihilation. We compute the energy deposition due to Coulomb scatterings of millicharged particles with protons in the  mantle and find that low-energy supernovae impose the most stringent constraints on millicharged particles in the mass range of $\sim(12 - 170)$ MeV. Furthermore, we find that the electron-positron annihilation process, previously omitted in supernova studies on millicharged particles, is the dominant production channel in the high-mass region. This leads to new constraints from both supernova cooling calculations and low-energy supernova analyses. We also investigate MCP production via processes involving thermal pions and find that these processes could dominate over electron-positron annihilation, albeit with significant uncertainties.

}

\begin{document}

\maketitle
\flushbottom

\section{Introduction}

Recently, 
there has been widespread interest in searches 
for new light particles beyond the standard model 
(SM) 
\cite{Alexander:2016aln}. 
One intriguing class of such particles is 
millicharged particles (MCPs),  
which possess a small electric charge
\cite{Jaeckel:2010ni, Fabbrichesi:2020wbt}.  
While terrestrial experiments provide leading constraints for high-mass MCPs  
\cite{Prinz:1998ua, Magill:2018tbb, ArgoNeuT:2019ckq, Marocco:2020dqu, Ball:2020dnx, 
Plestid:2020kdm, Kachelriess:2021man, ArguellesDelgado:2021lek, 
Du:2022hms, SENSEI:2023gie, Wu:2024iqm}, 
the low-mass range is best probed in celestial objects 
\cite{Dobroliubov:1989mr, Raffelt:1996wa,Davidson:2000hf, Vinyoles:2015khy,Chang:2018rso,  Fung:2023euv}, 
among which, supernovae (SNe), 
which are high-temperature and high-density stellar explosions,  
provide the most stringent limits in the MeV range 
\cite{Davidson:2000hf,Chang:2018rso}.

Constraints on new particles from supernovae are primarily derived 
from the cooling argument, 
which demands that any novel energy loss mechanism must be less efficient 
than the standard neutrino processes \cite{Raffelt:1996wa}. 
Another critical constraint comes from the calorimetric limit, which 
requires that any additional energy transfer from the SN core 
to the mantle must not exceed the explosion energy 
\cite{Falk:1978kf,Sung:2019xie,Caputo:2022mah}. 
Thus, it is of great interest to analyze
constraints on new particles from low-energy supernovae (LESNe), 
the underluminous Type II-P SNe that are 
10 to 100 times dimmer than the typical core-collapse SNe 
and can have an explosion energy as low as 0.1 
Bethe (B), where $1\, \rm B=10^{51}$ erg \cite{Caputo:2022mah}.  
LESNe have been observed in a number of SN events  
\cite{Chugai:1999en,Pastorello:2003tc,Pastorello:2009pt,10.1093/mnras/stu156,Pejcha:2015pca,Yang:2015ooa,Pumo:2016nsy,Murphy:2019eyu,Burrows:2020qrp,Yang:2021fka,
Teja:2024cht}, 
and are also confirmed in SN simulations 
\cite{Kitaura:2005bt,Fischer:2009af,Melson:2015tia,Radice:2017ykv,Lisakov:2017uue,Muller:2018utr,Burrows:2019rtd,Stockinger:2020hse,Zha:2021bev}. 
The calorimetric limits derived from LESNe can impose the most stringent 
constraints on various decaying or annihilating particles, 
including scalars \cite{Caputo:2022mah, Lella:2024dmx, Alda:2024cxn}, 
sterile neutrinos \cite{Chauhan:2023sci, Chauhan:2024nfa, Carenza:2023old}, 
and hidden sector fermions with significant self-interaction \cite{Fiorillo:2024upk}.

\begin{figure}[htbp]
    \centering
    \includegraphics[width=0.3\linewidth]{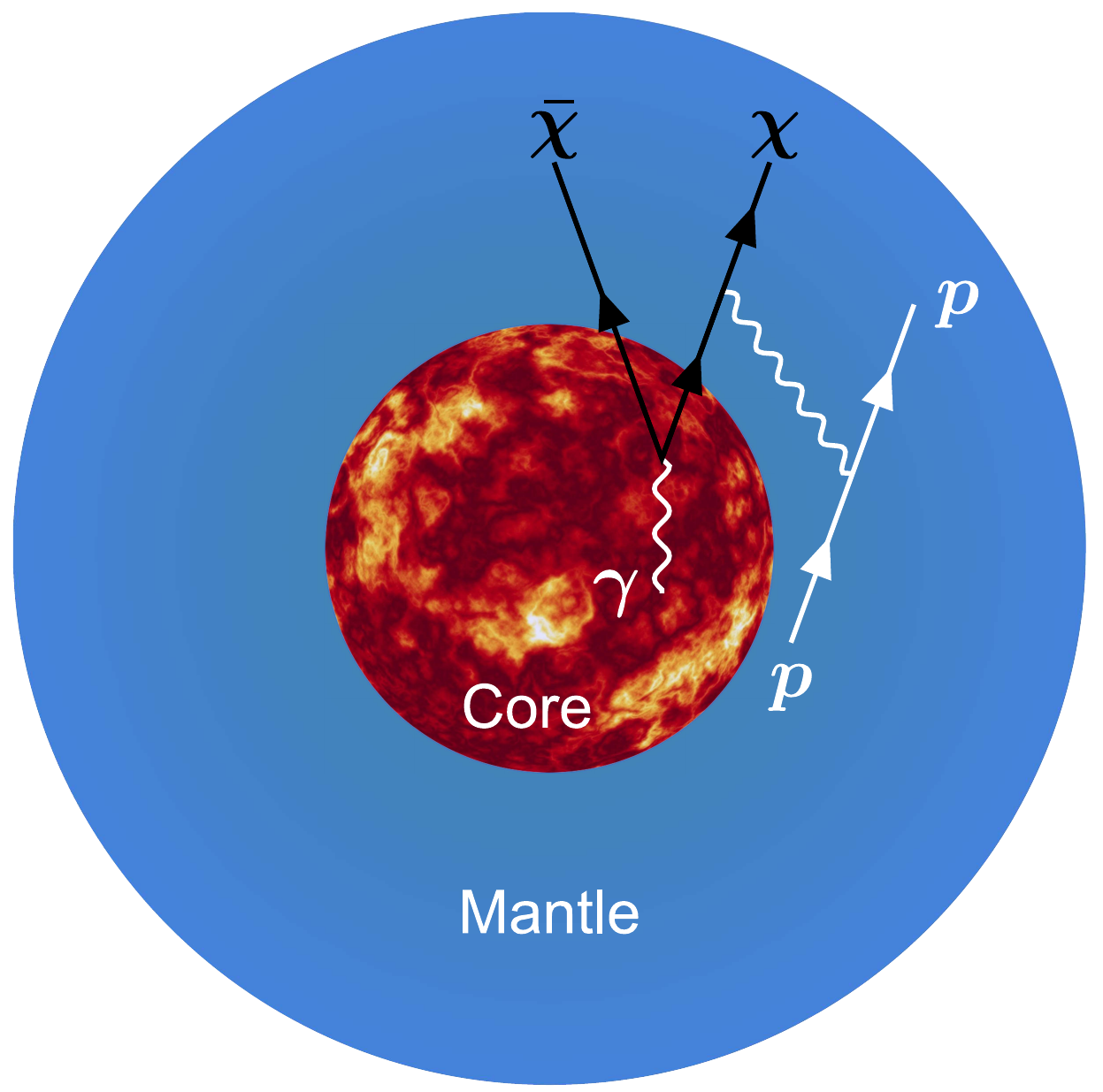}
    \caption{Schematic plot of the energy deposition in the mantle 
    from millicharged particles produced in the SN core.}
    \label{fig:SchematicDiagram}
\end{figure}

In this paper we study LESN constraints on MCPs. 
Unlike previous studies where 
the deposit energy in the mantle is due to decay or annihilation of new particles, 
MCPs deposit their kinetic energy in the 
mantle via Coulomb scattering with SM particles.
A schematic plot of the energy deposition in the mantle 
from MCPs produced in the SN core is shown in 
Fig.~\ref{fig:SchematicDiagram}. 
In our analysis we consider the minimal MCP model 
where the MCP $\chi$ is the only new particle beyond the SM and  
the interaction Lagrangian between the $\chi$ 
and the SM photon $A_{\mu}$ is 
\begin{equation}
\mathcal{L} = \epsilon e A_{\mu}\bar{\chi}\gamma^{\mu}\chi, 
\label{eq:model}
\end{equation}
where $e$ is the QED coupling constant, 
and $\epsilon$ is the millicharge of $\chi$.

MCPs can naturally arise in hidden $U(1)_X$ models, 
through either a kinetic mixing term 
\cite{Holdom:1985ag,Holdom:1986eq,Foot:1991kb} 
or a Stueckelberg mass mixing term 
\cite{Kors:2004dx,Cheung:2007ut,Feldman:2007wj} 
between the SM hypercharge gauge boson 
and the new $U(1)_X$ boson. 
In the case of kinetic mixing, 
the MCP $\chi$ must be accompanied by a massless $U(1)_X$ boson;  
however, for the mass mixing case, 
both $\chi$ and the $U(1)_X$ boson can be massive \cite{Feldman:2007wj}. 
Therefore, 
the minimal MCP model in Eq.~\eqref{eq:model}, 
commonly used in 
MCP phenomenological studies, 
can be regarded as the low-energy effective theory 
of the mass mixing case with a very heavy $U(1)_X$ boson. 
We note that the recent LESN analysis 
on self-interacting millicharged particles 
in Ref.~\cite{Fiorillo:2024upk} 
considered the kinetic mixing case with a massless $U(1)_X$ boson. 
The minimal MCP model used in our analysis 
differs significantly from that in 
Ref.~\cite{Fiorillo:2024upk}, 
as the latter contains an additional 
dark photon
in the hidden sector, 
leading to substantially different 
phenomenological implications compared to ours.

MCPs can be produced in the SN core via a number 
of processes, including 
plasmon decay, 
proton bremsstrahlung, and 
electron-positron annihilation, 
as shown in Fig.~\ref{fig:ThreeProcesses}. 
While previous studies have investigated only the first two processes 
\cite{Davidson:2000hf,Chang:2018rso}, 
we find that the electron-positron annihilation process  
is the dominant production channel of MCPs with large mass.
By considering all the three channels, 
we find that LESNe provide leading constraints 
on MCPs with mass $\sim(12-170)$ MeV, 
surpassing the energy loss limit from SN1987A \cite{Chang:2018rso}.

\section{Plasmon decay}

\begin{figure*}[htbp]
\centering
\includegraphics[width=0.3 \textwidth]{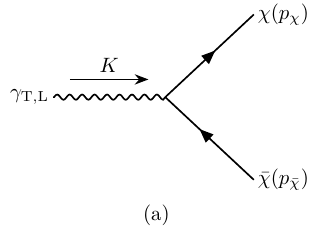}
\hspace{0.015 \textwidth}
\includegraphics[width=0.3 \textwidth]{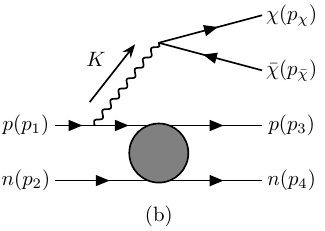}
\hspace{0.03 \textwidth}
\includegraphics[width=0.3 \textwidth]{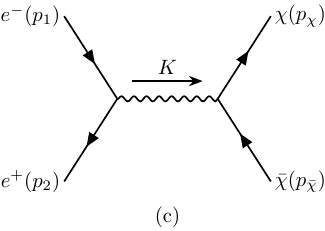}
  \caption{MCP production processes in the SN core: 
  (a) plasmon decay, 
  (b) proton bremsstrahlung, 
  and (c) electron-positron annihilation.} 
  \label{fig:ThreeProcesses}
\end{figure*}

We 
first discuss the plasmon decay process 
for the MCP production inside the SN core. 
Due to plasma effects, 
the decay process of $\gamma \to \chi\bar\chi$, 
which is forbidden in vacuum, can occur abundantly 
in the SN core. 
This is known as the plasmon decay process, 
as shown in Fig.~\ref{fig:ThreeProcesses}~(a).  
In Lorenz guage,
the decay widths of the longitudinal and the transverse photons into 
a pair of MCPs in the SN rest frame are given by 
\begin{equation}
\Gamma_{a}   = 
Z_{a} \frac{\epsilon^2 \alpha K^2 }{3 \omega_{a}}
f\left( \frac{m_\chi^2}{K^2} \right), 
\label{eq:plasmon:decay:TL}
\end{equation}
where $K^\mu =(\omega, {\bf k})$ is the photon momentum, 
$\alpha=e^2/4\pi$, 
$f\left( x \right) \equiv \sqrt{1-4x}\left(1+2x\right)$, 
$a = T$ and $a=L$ 
denote the transverse and longitudinal polarizations, respectively, and
$Z_a$ are the wave function renormalization factors 
\cite{Braaten:1993jw, Raffelt:1996wa}; 
see appendix \ref{appendix:plasmon:decay} 
for the detailed calculations.

For the LESN analysis, we adopt the one-zone model  
\cite{Caputo:2022mah} for the core, 
where 
the radius is $R_c = 12.9$ km, 
the temperature is $T_c = 30$ MeV, 
the nuclear density is $\rho_c = 3\times 10^{14}$ g/cm$^{3}$, and 
the proton abundance is $Y_p=0.15$.

In the relativistic limit, 
the production rate of MCPs per unit volume per unit energy in the 
core is given by 
\begin{equation}
\frac{d\Phi_a}{dE_\chi} = \frac{g_a}{2\pi^2} 
\int_0^\infty d k \, k^2\frac{\Gamma_a}{e^{\omega_a/T_c}-1} 
g(E_\chi,m_\chi,K),
\label{eq:plasmon:decay:flux}    
\end{equation}
where 
$k \equiv |{\bf k}|$,
$g_{L}=1$, 
$g_{T}=2$, and 
\begin{equation}
g(E_\chi,m_\chi,K) = 2\frac{\Theta(E_\chi-E_{\chi}^{-})\Theta(E_{\chi}^{+}-E_\chi)}{E_{\chi}^{+}-E_{\chi}^{-}}, 
\label{eq:chi:energy:spectrum}
\end{equation}
is the MCP energy spectrum in the  
SN
rest frame 
with $E_{\chi}^{\pm}= 
 \left( \omega \pm {k}
\sqrt{1-{4m_\chi^2}/{K^2}}
\right)/2$ 
being the maximal/minimal energy, 
and the factor of 2 account for the fact 
that there are two MCPs per decay.
See appendix \ref{appendix:plasmon:decay}
for the detailed calculations.

Plasmon decay is the dominant production channel of low-mass MCPs in the SN core. 
However, as the MCP mass increases so that it exceeds $\sqrt{K^2}/2$, 
the plasmon decay process ceases to occur. 
In the one-zone model, 
because the maximum value of 
the transverse photon mass is $\simeq 12$ MeV, 
the plasmon decay process can only produce 
MCPs with mass $\lesssim 6$ MeV. 
The proton bremsstrahlung and electron-positron annihilation processes 
become important for high-mass MCPs, 
because the typical nucleon (electron) energy is 
$\sim 45$ ($160$) MeV in the one-zone model.

\section{Proton bremsstrahlung}

We 
next discuss the proton bremsstrahlung process 
for the MCP production, 
as shown in Fig.~\ref{fig:ThreeProcesses} (b). 
We compute the differential cross section of 
the $np\to np\chi\bar{\chi}$ process via 
\cite{Gninenko:2018ter,Liang:2021kgw,Du:2022hms}  
\begin{equation}
\frac{d\sigma (np\to np\chi\bar{\chi})}{dK^2d\omega}
=
\frac{\epsilon^2\alpha}{3\pi}
\frac{1}{K^2}
\frac{d\sigma (np\to np\gamma)}{d\omega}
f\left(\frac{m_\chi^2}{K^2}\right), 
\label{eq:brem:differentialxsec}
\end{equation}
where 
${d\sigma (np\to np\gamma)}/{d\omega}$ is the 
differential cross section for emitting 
an off-shell photon with the momentum of 
$K^\mu = (\omega, {\bf k})$.  
Note that we do not include plasma effects for the 
photon propagator in the proton bremsstrahlung process to avoid 
double-counting with the plasmon decay process \cite{Chu:2019rok}.

In the one-zone model, 
the energy in the CM frame 
$E_{\rm cm} = ({\bf p}_1-{\bf p}_2)^2/4 m_N$ 
is about 90 MeV where $m_N$ is the nucleon mass. 
Note that for collisions with $E_{\rm cm} \approx 100$ MeV, 
the soft radiation approximation (SRA)  
agrees well with data 
for $\omega \ll E_{\rm cm}$, 
and underestimates the cross section by a factor of about 2 
for $\omega \approx E_{\rm cm}$ 
\cite{Huisman:1999zz, Huisman:1999ucq, Safkan:2007iy, Rrapaj:2015wgs}. 
Thus, we 
follow Refs.~\cite{Rrapaj:2015wgs,Chu:2019rok} 
to use the SRA to compute
$\sigma (np\to np\gamma)$: 
\begin{equation}
\frac{d\sigma (np\to np\gamma)}{d\omega}
= 
\sigma^T_{np}
\frac{d{\cal P}}{d\omega}, 
\label{eq:softphoton}
\end{equation}
where 
$\sigma^T_{np}$ 
is the transport cross section 
of the $np\to np$ process, and 
$d{\cal P}/d\omega$ is the photon splitting kernel. 
We use data from 
figure 3 of Ref.~\cite{Rrapaj:2015wgs} and 
figure 2 of Ref.~\cite{Brown:2018jhj} for 
$\sigma^T_{np}$.   
In the SRA, $d\mathcal{P}/d\omega$ is given by~\cite{Rrapaj:2015wgs,Chu:2019rok}
\begin{align} 
\frac{d{\cal P}}{d\omega} = 
\frac{4\alpha}{3\pi\omega}
\frac{E_{\rm cm}}{
m_N
}
f\left(\frac{K^2}{4\omega^2}\right). 
\label{eq:dif:splittingkernel}
\end{align}

The production rate of MCPs per unit volume per unit energy 
in this process is given by 
\begin{align}
\frac{d\Phi_{\rm pb}}{dE_\chi} = &
\frac{4 n_1n_2\epsilon^2\alpha}{3 \sqrt{m_N \pi^3 T_c^3}}
\int_{2m_{\chi}}^{\infty}dE_{\rm cm}E_{\rm cm}
e^{-{E_{\rm cm}}/T_c}\sigma^{{T}}_{np}(E_{\rm cm})
\nonumber \\ & \times 
\int_{4m_{\chi}^2}^{E_{\rm cm}^2} 
\frac{dK^2}{K^2} 
f\left(\frac{m_\chi^2}{K^2}\right)
\int_{\sqrt{K^2}}^{E_{\rm cm}} d\omega 
\frac{d\mathcal{P}}{d\omega}g(E_\chi,m_\chi,K). 
\label{eq:flux:pb}
\end{align}
Note that 
Eq.~\eqref{eq:flux:pb} has 
an extra factor of 4 compared to 
Ref.~\cite{Chu:2019rok}. 
See appendix \ref{appendix:proton:bremsstrahlung}
for the detailed calculations.

\section{Electron-positron annihilation}

The 
electron-positron annihilation process is another 
important MCP production channel, as 
shown in Fig.~\ref{fig:ThreeProcesses}~(c). 
The total electron-positron annihilation cross section 
is given by 
$\sigma_\mathrm{ann}=\sigma_T + \sigma_L$, where 
$\sigma_T$ ($\sigma_L$) is the cross section   
due to transverse (longitudinal) photons: 
\begin{equation}
\sigma_a = 
\frac{2\pi\epsilon^2\alpha^2}
{3\beta_e} 
\frac{N_a K^2 f\left({m_\chi^2}/{K^2}\right)}{(K^2-\mathrm{Re}\Pi_{a})^2+(\mathrm{Im}\Pi_a)^2}, 
\label{eq:eeann:xsec}
\end{equation}
where 
$\beta_e=\sqrt{1-4m_e^2/K^2}$,
Re$\Pi_a$ (Im$\Pi_a$) is the real (imaginary) part  
of the electromagnetic (EM) polarization tensor,  
$N_L = 1-E_-^2/(E_+^2-K^2)$, and 
$N_T = 1+4m_e^2/K^2+E_-^2/(E_+^2-K^2)$,  
where $E_\pm \equiv E_1 \pm E_2$ with 
$E_1$ ($E_2$) being the energy of the initial 
electron (positron). 
The electron-positron annihilation 
always occurs 
at $\sqrt{K^2}$ larger than the effective photon mass, 
due to the plasma effects. 
For example, in the one-zone model, 
the maximum value of the transverse photon mass is $\sim 12$ MeV,
which is smaller than $2m_e$, where $m_e \sim 9$ MeV,
the effective electron mass.

We use 
on-shell dispersion relations 
given in Ref.~\cite{Braaten:1993jw} 
to compute Re$\Pi_a$ in the off-shell region 
for the electron-positron annihilation process. 
This is because in the one-zone model, 
electrons are relativistic,  
and in the relativistic limit, Re$\Pi_a$ computed
with on-shell dispersion relations 
agrees well with the full analysis \cite{Scherer:2024uui}. 
The dominant contributions to Im$\Pi_a$ come from 
the proton bremsstrahlung process and its inverse process. 
We find that Im$\Pi_a$ at most  
contributes $\sim 2$\% to the total cross section. 
We thus neglect Im$\Pi_a$ in our analysis.
See appendix \ref{appendix:plasma:lorenz}
for the detailed analysis on 
the contributions from both Re$\Pi_a$ and Im$\Pi_a$ 
to the electron-positron annihilation cross section.

The production rate of MCPs per unit volume per unit energy 
due to electron-positron annihilation is given by
\begin{align}
    \frac{d\Phi_{\rm ann}}{dE_\chi}  = &
    \frac{1}{16\pi^4}
    \int_{4m_{\mathrm{th}}^2}^{\infty} dK^2 K^2 \beta_e
    \int_{\sqrt{K^2}}^{\infty}dE_+\int_{-E_-^m}^{E_-^m} dE_-  
    f_{1} (E_1) f_{2} (E_2) \, 
    \sigma_{\mathrm{ann}} \, 
    g(E_\chi,m_\chi,K),
\label{eq:phi:ee:2}
\end{align}
where 
$E_-^m \equiv \beta_e \sqrt{E_+^2-K^2}$,  
$m_{\mathrm{th}}\equiv\max\{m_e,m_{\chi}\}$, and 
$f_1(E_1)$ and $f_2(E_2)$ are the Fermi-Dirac distributions 
for the initial electron and positron, respectively. 
The electron chemical potential in $f_1(E_1)$ and $f_2(E_2)$  
is $\mu \simeq 167$ MeV.
See appendix \ref{appendix:electron:positron:annihilation} 
for the detailed calculations.

\section{Energy deposition}

\begin{figure}[htbp]
\centering
\includegraphics[width=0.35\textwidth]{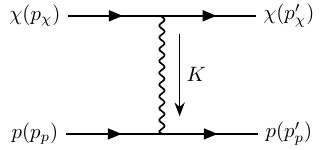}
  \caption{Energy deposition of MCPs via the Coulomb scattering with protons in the mantle.}
  \label{fig:chipScattering}
\end{figure}

The 
energy transfer from the core to the mantle, mediated 
by MCPs, 
primarily occurs 
through the Coulomb scattering of MCPs with protons \cite{Davidson:2000hf}, 
as shown in Fig.~\ref{fig:chipScattering}.

We compute the energy loss of $\chi$ per unit length via 
\begin{equation}
\frac{dE_{\chi}}{d x}=- n_p \int dE_R \frac{d\sigma_{\chi p}}{dE_R} E_R,      
\end{equation}
where $n_p$ is the proton number density in the mantle, 
${d\sigma_{\chi p}}/{dE_R}$ is the differential Coulomb scattering 
cross section, and 
$E_R$ is the recoil energy received by protons 
in the mantle. 
For the 2-to-2 elastic scattering, the recoil energy is 
$E_R = E_R^{\rm max} (1-\cos\theta)/2$, 
where $\theta$ is the scattering angle in the center-of-mass frame, 
and $E_R^{\rm max}$ is the maximum recoil energy. 
Thus, one has 
\begin{equation}
\frac{dE_{\chi}}{d x}=- \frac{1}{2} n_p E_R^{\rm max} \sigma_{\chi p}^T, 
\end{equation}
where $\sigma_{\chi p}^T = \int d\Omega 
(d\sigma_{\chi p}/d\Omega) (1-\cos\theta)$ 
is the transport cross section. 
In the plasma, one has to take into account the 
Debye screening effects, which can be achieved by multiplying 
the scattering cross section with the  
${K^2}/{(K^2+k_D^2)}$ factor, leading to 
\cite{Davidson:2000hf} 
\begin{equation}
\sigma_{\chi p}^T =
\frac{2\pi\epsilon^2\alpha^2}{E_\chi^2}
\left[\frac{2+z}{2}\ln\left(\frac{2+z}{z}\right)-1\right],
\label{eq:sigma:transport}
\end{equation}
where $K$ is the four-momentum of the exchanged photon, 
$k_D = 2\sqrt{\pi \alpha n_p/T}$ is the Debye scale, and 
$z = {k^2_D}/{2E_\chi^2}$. 
Since the mantle is colder than the SN core, for simplicity, 
we assume that protons are initially at rest; 
in this case, one has 
\begin{equation}
E_R^{\rm max} =  
\frac{2 m_p (E_\chi^2-m_\chi^2)}
{m_p^2 + m_\chi^2 + 2 m_p E_\chi}. 
\label{eq:ER:Max}
\end{equation} 
The energy deposited by a single $\chi$ particle
in the mantle is given by
\begin{equation}
\Delta E_\chi = \frac{1}{2} \int dx 
 n_p E_R^{\rm max} \sigma_{\chi p}^T, 
 \label{eq:E:loss}
\end{equation}
where the distance traversed in the mantle is three light-seconds, and, 
for simplicity, 
we have taken the integral along the radial direction. 
According to the Garching group's model SFHo-18.8
\cite{Bollig:2020xdr, CCSNarchive}, 
the mass inside the neutrinosphere 
is about the same as that of the collapsed neutron star. 
Thus only the energy deposited outside the neutrinosphere by MCPs 
is included in the calculation of the explosion energy.

In our analysis 
we adopt the following profiles 
for the mass density and temperature of the SN mantle
\cite{Chang:2016ntp,Raffelt:1996wa}:  
\begin{align}
\rho(r)&=\rho_{c}\times(r/R_{c})^{-\nu},\quad r>R_{c}
\label{eq:SN:profile:power:law:rho},\\
T(r)&=T_{c}\times(r/R_{c})^{-\nu/3},\quad r>R_{c},
\label{eq:SN:profile:power:law:T}
\end{align}
where $\nu=5$. 
Following Ref.~\cite{Cooperstein:1988fz}, 
we determine the neutrinosphere at the mass density of  
$\rho \simeq 10^{12}\ \mathrm{g/cm}^3$, 
which then leads to $R_\nu\simeq40$ km. 
We then use $Y_p=0.15$ to compute the proton number density 
in the radial range of interest. 
This value of the proton abundance 
is supported by the Garching group's model SFHo-18.8 
\cite{Bollig:2020xdr, CCSNarchive} which has  
$0.15 \lesssim Y_p \lesssim 0.53$ for  
$40 \lesssim r \lesssim 80$ km.

\section{LESN constraints on MCPs}

\begin{figure}[htbp]
\centering
\includegraphics[width=0.5 \textwidth]{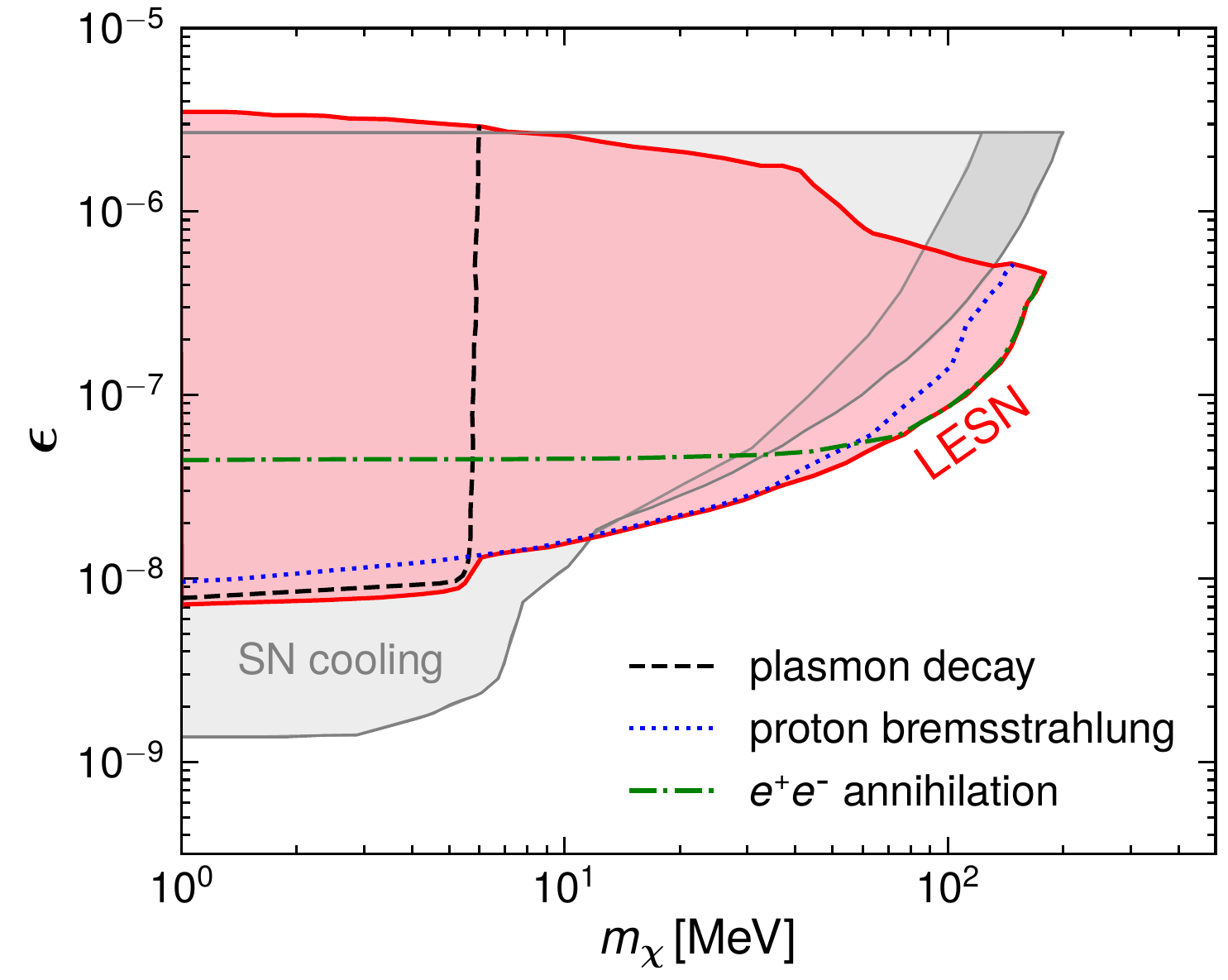} 
\caption{LESN constraints on MCPs (red), where 
the one-zone model \cite{Caputo:2022mah} for the SN core is used 
and the energy deposited outside the neutrinosphere is 
constrained to be $<0.1$ B. 
Also shown are the lower boundaries of the exclusion region 
using a single production channel:
plasmon decay (black-dashed), 
proton bremsstrahlung (blue-dotted), 
and electron-positron annihilation (green-dotdashed) processes. 
The light-gray and dark-gray regions indicate the SN cooling limits    
where the $e^+e^-$ annihilation is omitted \cite{Chang:2018rso} and included, respectively.
} 

\label{fig:main:result:1}
\end{figure}

We 
compute the total energy deposition in the mantle 
due to MCPs via 
\begin{equation}
 E_{m} =  
{\rm lapse^2 }
\times 
4\pi \Delta t  \int_0^{R_c}  
dr r^2\int_{m_\chi'}^\infty dE_\chi\frac{d\Phi}{dE_\chi} \Delta E_\chi,  
\label{eq:energy:deposit}
\end{equation}
where 
$\Delta t = 3$ s, 
$\Delta E_\chi$ is given in Eq.~\eqref{eq:E:loss}, 
$d\Phi/dE_\chi$ is the total MCP flux, 
${\rm lapse} \equiv \sqrt{1-2GM/R_c}$ 
\cite{Caputo:2022mah, Chauhan:2023sci, Chauhan:2024nfa}, 
and $m_\chi'=m_\chi/{\rm lapse}$. 
The $\mathrm{lapse}^2$ factor accounts for 
the gravitational redshift effect since the energy 
constraint is measured at Earth  
\cite{Caputo:2022mah, Chauhan:2023sci, Chauhan:2024nfa}.  
The total MCP flux includes contributions from 
plasmon decay, 
proton bremsstrahlung, 
and electron-positron annihilation processes.
To compute the LESN constraints on MCPs, 
we require the total energy deposition in the 
mantle to satisfy $E_m \leq 0.1$ B. 
Fig.~\ref{fig:main:result:1} shows 
that the LESN constraints probe new parameters 
in the mass range of $\gtrsim 12$ MeV, 
surpassing previous SN cooling limits given in 
Refs.~\cite{Chang:2018rso,Davidson:2000hf}.

Fig.~\ref{fig:main:result:1} shows the dominant 
production channel for different MCP masses.
In the low-mass region of $m_\chi \lesssim 6~\mathrm{MeV}$, 
the production of MCPs in the SN core 
is dominated by the plasmon decay process \cite{Davidson:2000hf}. 
For higher MCP masses, the proton bremsstrahlung and 
electron-positron annihilation processes become more important. 
Owing to the high number density of nucleons in the SN core, 
the proton bremsstrahlung process is the 
main MCP production channel in the mass range 
of $6~\mathrm{MeV}\lesssim m_\chi \lesssim 57~\mathrm{MeV}$. 
In the mass region of $m_\chi \gtrsim 57~\mathrm{MeV}$, 
the electron-positron annihilation process becomes the most 
important MCP production channel. 
This is partly because, in the SN core, the average energy of 
the electrons 
($\sim 160$ MeV)
is much higher than the average kinetic energy of the nucleons 
($\sim 45$ MeV).
We note that the electron-positron annihilation process is often omitted 
in previous studies due to the small number density of the positron.  
However, the high temperature $T\gg m_e$ in the LESN core 
actually leads to a substantial population of positrons, 
making the electron-positron annihilation the most important production channel 
for high-mass MCPs. 
We also analyze the SN cooling limits by taking into account 
contributions from the electron-positron annihilation process. 
See appendix \ref{appendix:SN:cooling:limit}
for detailed calculations. 
Fig.~\ref{fig:main:result:1} shows that 
incorporating the electron-positron annihilation channel 
strengthens the SN cooling limits, 
extending the constraint to higher masses.

For sufficiently large $\epsilon$, MCPs become effectively trapped, 
forming an MCP-sphere. In this regime, they cannot 
deposit significant energy outside the neutrinosphere 
(see appendix \ref{appendix:large:coupling}
for detailed calculations). 
Consequently, 
the LESN exclusion region exhibits an upper boundary, 
beyond which the parameter space is allowed, 
as shown in Fig.~\ref{fig:main:result:1}.

\section{Pionic process}\label{sec:PionicProcess}

Thermal pions can be significant in hot dense matter 
and can have great influence on SN physics \cite{Fore:2019wib}, 
such as axion emission 
\cite{Carenza:2020cis} 
(see also 
\cite{Fischer:2021jfm,Choi:2021ign,Lella:2022uwi,Lella:2023bfb,Carenza:2023lci,Cavan-Piton:2024ayu,Lella:2024dmx,Lella:2024hfk,Manzari:2024jns,Arias-Aragon:2024gdz,Alonso-Gonzalez:2024ems,Carenza:2025uib,Candon:2025fnb}), 
and dark gauge boson emission \cite{Shin:2022ulh}. 
Thus, it is of great interest to compare the MCP flux from pionic process
to the three processes shown in Fig.~\ref{fig:ThreeProcesses}.

\begin{figure}[htbp]
\centering
\includegraphics[width=\linewidth]{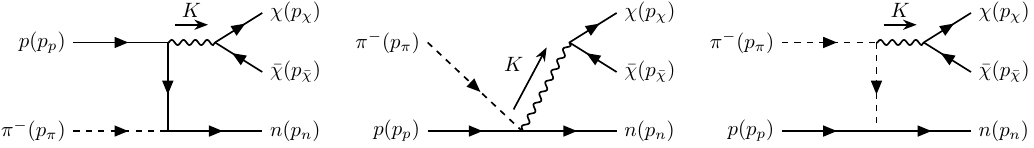}
\caption{Tree diagrams for the MCP production in the 
pionic process $\pi^-p\rightarrow n\chi\bar\chi$. 
These diagrams are obtained by attaching $\chi\bar\chi$ 
to the final state photon line in the diagrams in 
Fig.~1 of Ref.~\cite{Shin:2022ulh}.  
}
\label{fig:PionProcess}
\end{figure}

Fig.~\ref{fig:PionProcess} shows the three 
tree diagrams for the MCP production in 
pionic process, $\pi^-p\rightarrow n\chi\bar\chi$, 
where $\pi^-$ is the negatively charged pions; 
these diagrams are obtained by attaching $\chi\bar\chi$ 
to the final state photon line in the diagrams in 
Fig.~1 of Ref.~\cite{Shin:2022ulh}. 
The production rate of MCPs
per unit volume 
per unit energy
due to this pionic process
is
\begin{align}
\frac{d\Phi_\mathrm{pion}}{dE_\chi}
=&\frac{\epsilon^2e^4g_A^2}{96}
\sqrt{\frac{2(m_N^\ast)^3T_c^9}{\pi^{14}f_\pi^4}}z_pz_\pi
\int dx_p\frac{x_p^2}{e^{x_p^2}+z_p}\frac{e^{x_p^2}}{e^{x_p^2}+z_n}\int dx_\pi\frac{x_\pi^2}{e^{\kappa_\pi-y_\pi}}g(E_\chi,m_\chi,x_\pi)
\nonumber\\
&\times
\int_{4m_\chi^2}^{E_\pi^2}\frac{dK^2}{K^2}
f
\left(
\frac{m_\chi^2}{K^2}
\right)
\sqrt{1-\frac{K^2}{E_\pi^2}}\textbf{m}^2,
\label{eq:ProductionRate:Pion}
\end{align}
where
$f_\pi\simeq93$ MeV is the pion decay constant~\cite{Peskin:1995ev}, 
$g_A=1.2723$ is the axial-vector coupling constant
\cite{ParticleDataGroup:2024cfk},
$z_i=\exp[(\mu_i-m_i)/T_c]$
is the fugacity of a particle $i$,
$\textbf{m}^2=1+(\tilde{y}_\pi^2/\tilde{\kappa}_\pi^2)[\tilde{\kappa}_\pi^2/2\kappa_\pi^2+y_\pi^2\tilde{\kappa}_\pi^2/2(\tilde{\kappa}_\pi^4-x_\pi^2\kappa_\pi^2)-\mathrm{arctanh}(x_\pi\kappa_\pi/\tilde{\kappa}_\pi^2)/(x_\pi\kappa_\pi/\tilde{\kappa}_\pi^2)]$,
$x_p=|\textbf{p}_p|/\sqrt{2m_N^\ast T_c}$,
$x_\pi=|\textbf{p}_\pi|/T_c$,
$\kappa_\pi=\omega/T_c$,
$\tilde{\kappa}_\pi=\tilde{\omega}/T_c$,
$\tilde{\omega}=[(\omega^2+\textbf{p}_\pi^2+m_\pi^2)/2]^{1/2}$,
$y_\pi=m_\pi/T_c$, 
$\tilde{y}_\pi^2=p_\pi^2/T_c^2$, 
and $m_N^\ast$ is the Landau effective mass of nucleon. 
For the one-zone model where $\rho=3\times10^{14}\ \mathrm{g/cm}^3$, 
we use $m_N^\ast/m_N = 0.58$, which is adopted from 
Fig.~1 of Ref.~\cite{Carenza:2019pxu}. 
The chemical potential of $\pi^-$ is
$\mu_{\pi^-}=\mu_n-\mu_p$,
and 
the pion dispersion is
\begin{equation}
E_\pi=\sqrt{\textbf{p}_\pi^2+m_\pi^2}+\Sigma_\pi(|\textbf{p}_\pi|),
\end{equation}
where $\Sigma_\pi(|\textbf{p}_\pi|)$ is the real part of the momentum-dependent pion self-energy
\cite{Fore:2019wib}.
See Appendix \ref{appendix:PionicProcess}
for the detailed calculations
on Eq.~(\ref{eq:ProductionRate:Pion}).

\begin{figure}[htbp]
\centering
\includegraphics[width=0.5 \textwidth]{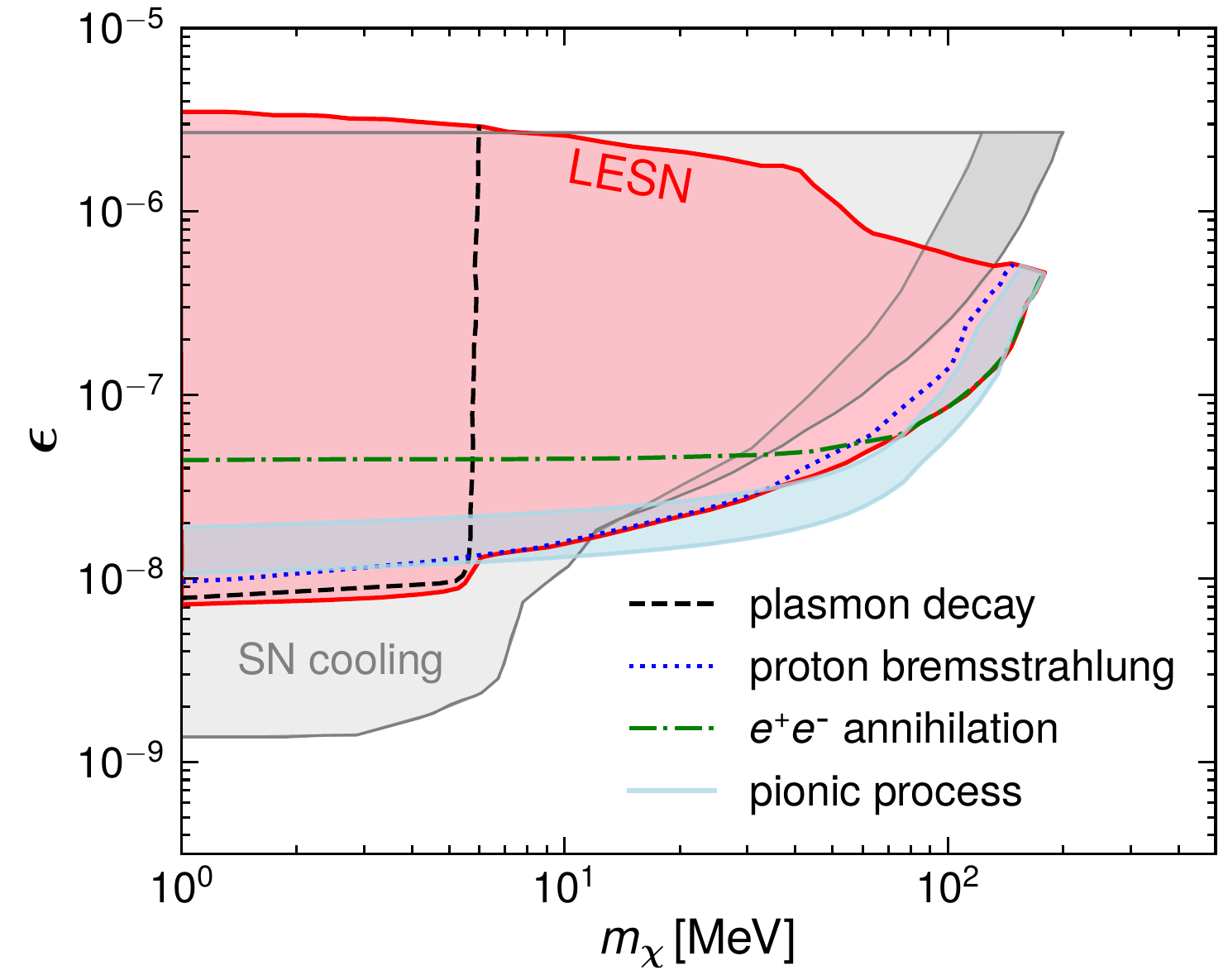} 
\caption{Same as Fig.~\ref{fig:main:result:1} except the blue band, 
which represents contributions from pionic processes, 
with $Y_{\pi^-}=10^{-3} \, (10^{-2})$  
for the upper (lower) boundary.
} 

\label{fig:main:result:2}
\end{figure}

We note that although thermal pions may be abundant in SN cores, 
accurately determining the $\pi^-$ abundance 
requires their inclusion in detailed SN simulations. 
The absence of such systematic treatments in the literature 
introduces uncertainty regarding the true extent of the enhancement 
that pionic processes 
may contribute to SN cooling bounds on new physics particles
\cite{Lella:2023bfb,Caputo:2024oqc,Harris:2024ssp}. 
We also note that the calculations of pion mass contain 
large uncertainties; for example, 
the pion mass is found to be in a wide range of 200--260 MeV 
for the baryon density of $n_B=0.16~{\rm fm}^{-3}$, 
as shown in Fig.~6 of Ref.~\cite{Fore:2023gwv}. 
To account for the uncertainty in the thermal pion distribution, 
we consider two different pion fractions, 
as suggested 
by recent studies 
\cite{Vijayan:2023qrt,Harris:2024ssp,Carenza:2020cis}: 
(1) $Y_{\pi^-} = 10^{-2}$ 
\cite{Carenza:2020cis}, 
and (2) $Y_{\pi^-} = 10^{-3}$ 
\cite{Harris:2024ssp}.

In Fig.~\ref{fig:main:result:2}, 
we compare the LESN constraints on MCPs 
from pionic processes (blue shaded band) 
with those from the three channels 
shown in Fig.~\ref{fig:ThreeProcesses} 
(red shaded region). 
The lower and upper boundaries of the blue band 
correspond to $Y_{\pi^-} = 10^{-2}$ and $Y_{\pi^-} = 10^{-3}$, respectively. 
For a pion abundance of $Y_{\pi^-} = 10^{-2}$, 
pionic production dominates over proton bremsstrahlung 
and $e^+e^-$ annihilation for MCP masses above 
$m_\chi \gtrsim 5$ MeV. 
However, 
for a lower pion abundance of $Y_{\pi^-} = 10^{-3}$, 
pionic production becomes subdominant compared to other production channels. 
We thus conclude that while thermal pions can 
significantly enhance the SN cooling bound, whether 
they provide the dominant contribution requires 
more careful investigation, given the current uncertainty.

\section{Conclusions}

In this paper 
we study LESN constraints on MCPs. 
We consider three MCP production channels in the SN core:  
plasmon decay, 
proton bremsstrahlung, and 
electron-positron annihilation.
We compute the energy deposited in the SN mantle 
via Coulomb scatterings with protons. 
By requiring the energy deposited in the mantle 
to be less than 0.1 B, 
we obtain new constraints on MCPs from LESNe.
We find that LESN constraints probe 
new MCP parameter space 
in the mass range of $\gtrsim 12$ MeV, 
surpassing previous SN cooling limits. 
Additionally, 
we find that the electron-positron annihilation process, 
previously overlooked in SN studies of MCPs, 
dominates production in the high-mass region; 
incorporating this channel strengthens both 
LESN and SN cooling limits.

\begin{acknowledgments}

We thank 
Hans-Thomas Janka and 
Yonglin Li  
for insightful discussions and correspondence.
We also appreciate   
Damiano F.G. Fiorillo 
and 
Edoardo Vitagliano 
for their valuable comments on our 
initial treatment of the LESN constraints.
The work is supported in part by the 
National Natural Science Foundation of China under Grant 
Nos.\ 12275128 and 12147103. 
The Feynman diagrams are created using the tikz-feynman package 
in \LaTeX\ \cite{Ellis:2016jkw}.
\end{acknowledgments}

\appendix

\section{Plasma effects in Lorenz gauge}
\label{appendix:plasma:lorenz}

In this section we provide a brief discussion 
on the plasma effects that are relevant for our analysis.
The properties of photons and electrons 
are significantly affected by the plasma effects in 
a supernova core, 
which are important for 
the production of millicharged particles. 
The leading order plasma effects on photons are 
encoded in the electromagnetic (EM) polarization tensor \cite{Braaten:1993jw}: 
\be
 {\rm Re} \Pi^{\mu\nu}=16\pi\alpha\int\frac{d^3 p}{(2\pi)^3}
 \frac{1}{2E}[f_{e^-}(E)+f_{e^+}(E)]
\frac{K\cdot P(P^\nu K^\mu+P^\mu K^\nu-P\cdot K g^{\mu\nu})-K^2 P^\mu P^\nu}{(K\cdot P)^2  {- (K^2)^2/4}},
\label{eq:pimunu}
\ee
where $ {K}^\mu=(\omega,\textbf{k})$ is the photon momentum, 
$ {P}^\mu=(E,\textbf{p})$ is the momentum of the electron (or positron), 
and 
$f_{e^-}(E) = (e^{(E-\mu)/T}+1)^{-1}$ and 
$f_{e^+}(E) = (e^{(E+\mu)/T}+1)^{-1}$ 
are the electron and positron distributions, respectively. 
Following Ref.~\cite{Braaten:1993jw}, we 
use $k \equiv |{\bf k}|$ 
and $p \equiv |{\bf p}|$ here to simplify the 
expressions.

In Lorenz gauge, the effective photon propagator in the plasma is  
\begin{equation}
\tilde D^{\mu\nu}(\omega,k) 
=
\sum_{a = \pm, L}
\frac{i}{ {K^2}
- {\rm Re} \Pi_{a}(\omega,k)
- i {{\rm Im} \Pi_{a}(\omega,k)}
}  \epsilon^\mu_{a}\epsilon^{\nu*}_{a},
\label{eq:lorenz:propagator}
\end{equation}
where 
$\epsilon_\pm^\mu = (0,1,\pm i,0)/\sqrt{2}$ 
and 
$\epsilon_L^\mu = (k,0,0,\omega)/\sqrt{K^2}$  
are the transverse and longitudinal polarization vectors, 
respectively.  
For the transverse polarizations, 
we have $\Pi_{T} \equiv \Pi_{+} = \Pi_{-}$.
In Lorenz gauge, 
$\Pi_L = ({K^2}/{k^2})\Pi^{00}$, and 
$\Pi_T = \Pi^{11} = \Pi^{22}$. 
The real parts of the polarization functions are given by  
\begin{align}
 {\rm Re}\Pi_L(\omega,k) & = 
\frac{4\alpha}{\pi}\frac{K^2}{k^2}\int_0^\infty 
 {dp} 
\frac{p^2}{E}\left(\frac{\omega}{vk}\ln\frac{\omega+vk}{\omega-vk}
-1-\frac{\omega^2-k^2}{\omega^2-v^2 k^2}\right)[f_{e^-}(E)+f_{e^+}(E)], 
\\ 
 {\rm Re}\Pi_T(\omega,k) & = \frac{4\alpha}{\pi}\int_0^\infty 
 {dp} 
\frac{p^2}{E}\left(\frac{\omega^2}{k^2}-\frac{\omega^2-k^2}{k^2}\frac{\omega}{2vk}\ln\frac{\omega+vk}{\omega-vk}\right)[f_{e^-}(E)+f_{e^+}(E)], 
\end{align}
where $v=p/E$. 
The dispersion relations are determined by 
$K^2 =  {\rm Re} \Pi_{a}$. 
The residues of the propagators at the poles then 
lead to effective photon polarization vectors of $\sqrt{Z_a}\epsilon_a^\mu$ 
where 
\begin{align} 
Z^{-1}_{a}(k) = 
1-\frac{\partial  {\rm Re}\Pi_{a}}{\partial\omega^2}(\omega_{a}(k),k).
\label{eq:ZL2}
\end{align}

\subsection{Relativistic limit}
\label{appendix:relativistic}

For the LESN core, we adopt the one-zone model, 
where the temperature is $T=30$ MeV, 
the nuclear density is 
$\rho=3\times 10^{14}$ g/cm$^{3}$, 
and the proton abundance is $Y_p=0.15$
\cite{Caputo:2022mah}. 
This leads to an electron chemical potential of 
$\mu \simeq 167$ MeV. 
Because both $T\gg m_e$ and $\mu \gg m_e$ are satisfied 
in the one-zone model, 
we use various expressions in the 
relativistic limit given in Ref.~\cite{Braaten:1993jw} for our analysis.

In the relativistic limit, the plasma frequency is given by 
\begin{equation}
\omega_p^2 = \frac{4\alpha}{3\pi} 
\left(
\mu^2 + \frac{1}{3} \pi^2 T^2
\right). 
\end{equation}
The real part of 
the polarization functions $ {\rm Re}\Pi_{L,T}$ 
in the relativistic limit 
are given by \cite{Braaten:1993jw, Raffelt:1996wa}
\begin{align}
 {\rm Re}\Pi_{L}(\omega,k) & = 3\omega_p^2\frac{K^2}{k^2}\left(\frac{\omega}{2k}\ln\frac{\omega+k}{\omega-k}-1\right),
\label{eq:longitudinal:polarization:function:B}\\
 {\rm Re}\Pi_{T}(\omega,k) & = \omega_p^2\frac{3\omega^2}{2k^2}\left(1-\frac{\omega^2-k^2}{\omega^2}\frac{\omega}{2k}\ln\frac{\omega+k}{\omega-k}\right). 
\label{eq:transverse:polarization:function:B}
\end{align}
The dispersion relations $\omega_{L,T}(k)$ 
in the relativistic limit are determined by 
the transcendental equations: 
\begin{align}
&\omega_L^2=\omega_p^2\frac{3\omega_L^2}{k^2}\left(\frac{\omega_L}{2k}
\ln\frac{\omega_L+k}{\omega_L-k}-1\right), 
\quad
0 \leq k <\infty,
\\
&\omega_T^2=k^2+\omega_p^2\frac{3\omega_T^2}{2k^2}\left(1-\frac{\omega_T^2-k^2}{\omega_T^2}\frac{\omega_T}{2k}\ln\frac{\omega_T+k}{\omega_T-k}\right),
\quad 
0\leq k<\infty. 
\end{align}
The residue factors in the relativistic limit are 
\cite{Raffelt:1996wa}
\begin{align}
&Z_L=\frac{2(\omega_L^2-k^2)}{3\omega_p^2-(\omega_L^2-k^2)}\frac{\omega_L^2}{\omega_L^2-k^2},
\label{eq:ZL:r:limit}
\\&Z_T=\frac{2\omega_T^2(\omega_T^2-k^2)}{\omega_T^2[3\omega_p^2-2(\omega_T^2-k^2)]+(\omega_T^2+k^2)(\omega_T^2-k^2)}.
\end{align}

\begin{figure}[htbp]
\centering
\includegraphics[width=0.4 \textwidth]{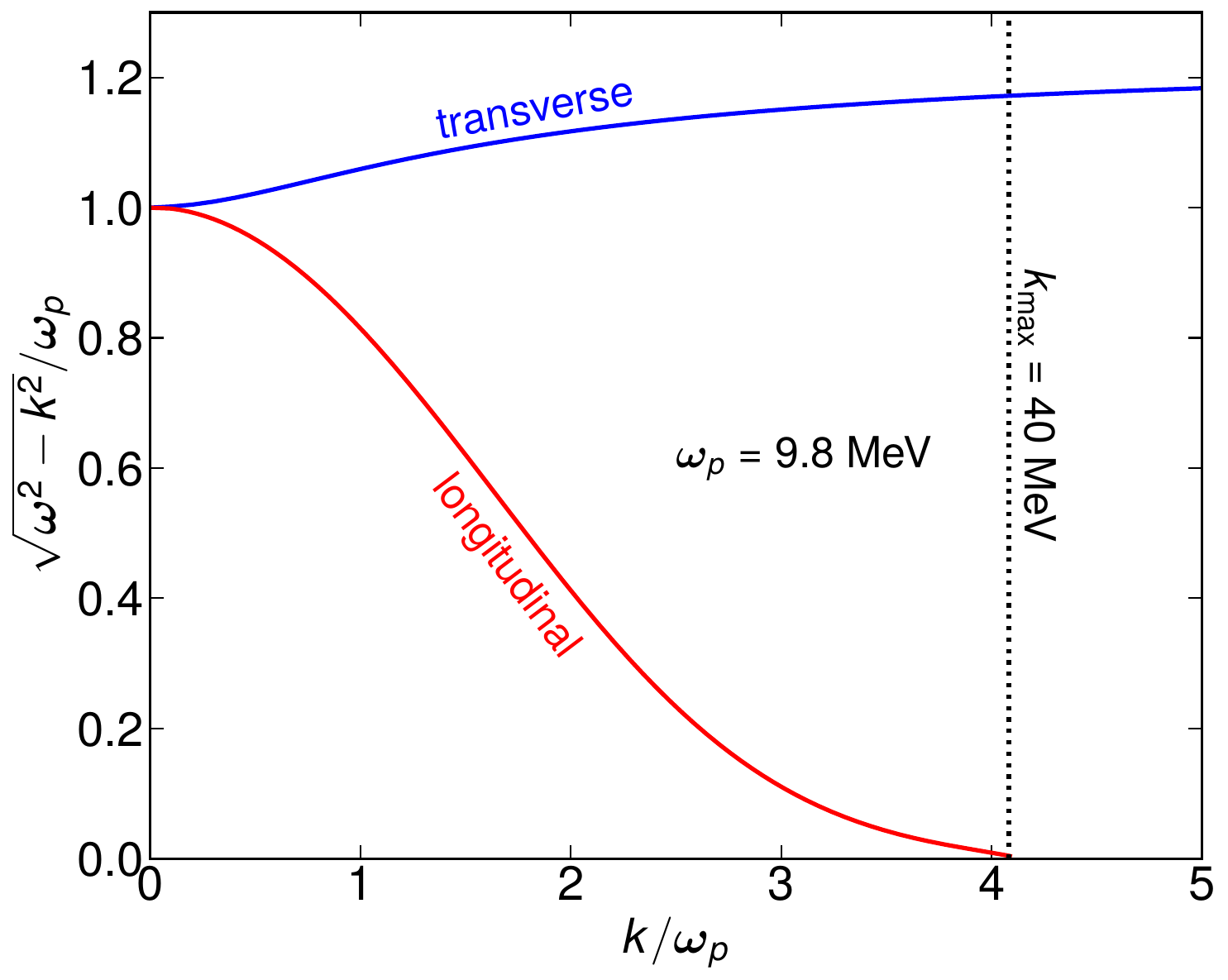} 
  \caption{The ratio of the photon mass to  the plasma frequency for the transverse (blue) and longitudinal (red) photons, as a function of the ratio of the photon momentum to the plasma frequency, 
  in the one-zone model  \cite{Caputo:2022mah}, where $\omega_p \simeq 9.8$ MeV, 
  and $k_{\rm max} \simeq 40$ MeV for the longitudinal photon.}
  \label{fig:ph:plasma:mass}
\end{figure}

By using the one-zone parameters,  
we obtain $\omega_p \simeq 9.8$ MeV. 
In Fig.~\ref{fig:ph:plasma:mass}, 
we compute
the masses of the transverse and longitudinal photons  
as a function of the photon momentum 
for the one-zone model 
in the relativistic limit.

\subsection{On-shell approximation in the off-shell region}
\label{appendix:off:shell}

In plasma, electrons (and positrons) acquire an in-medium mass so that 
electrons and positrons annihilate through   
$s$-channel intermediate photons with $\sqrt{K^2}$ larger than 
photon masses. 
For example, in the one-zone model, 
the effective electron mass is $\sim 9$ MeV, 
and the maximum value of the transverse photon mass is $\sim 12$ MeV. 
It is thus of importance to accurately determine both the real part 
and the imaginary part of $\Pi$ in the off-shell region of the photon.

\begin{figure}[htbp]
\centering
\includegraphics[width=0.45 \textwidth]{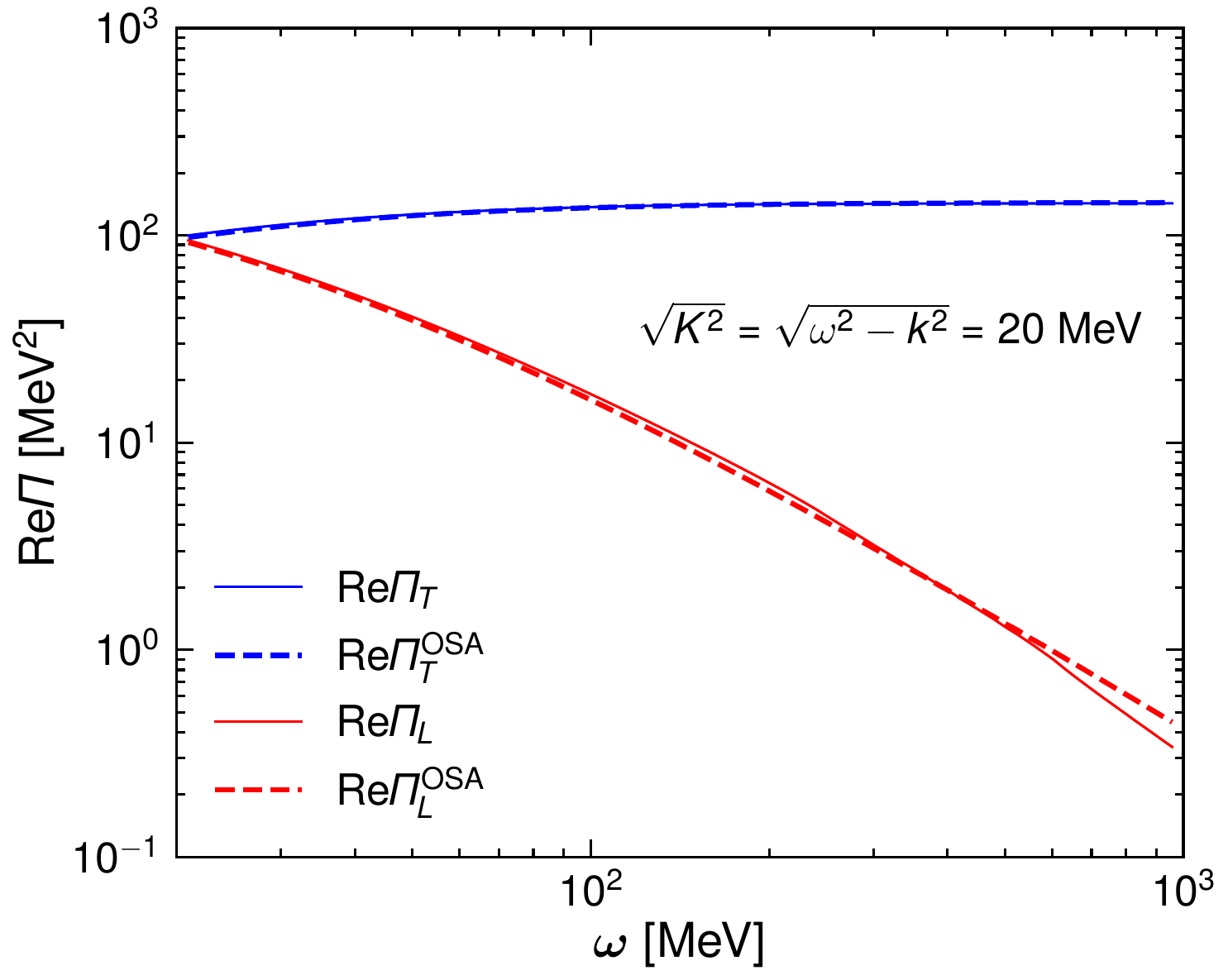} 
\hspace{0.5 cm}
\includegraphics[width=0.45 \textwidth]{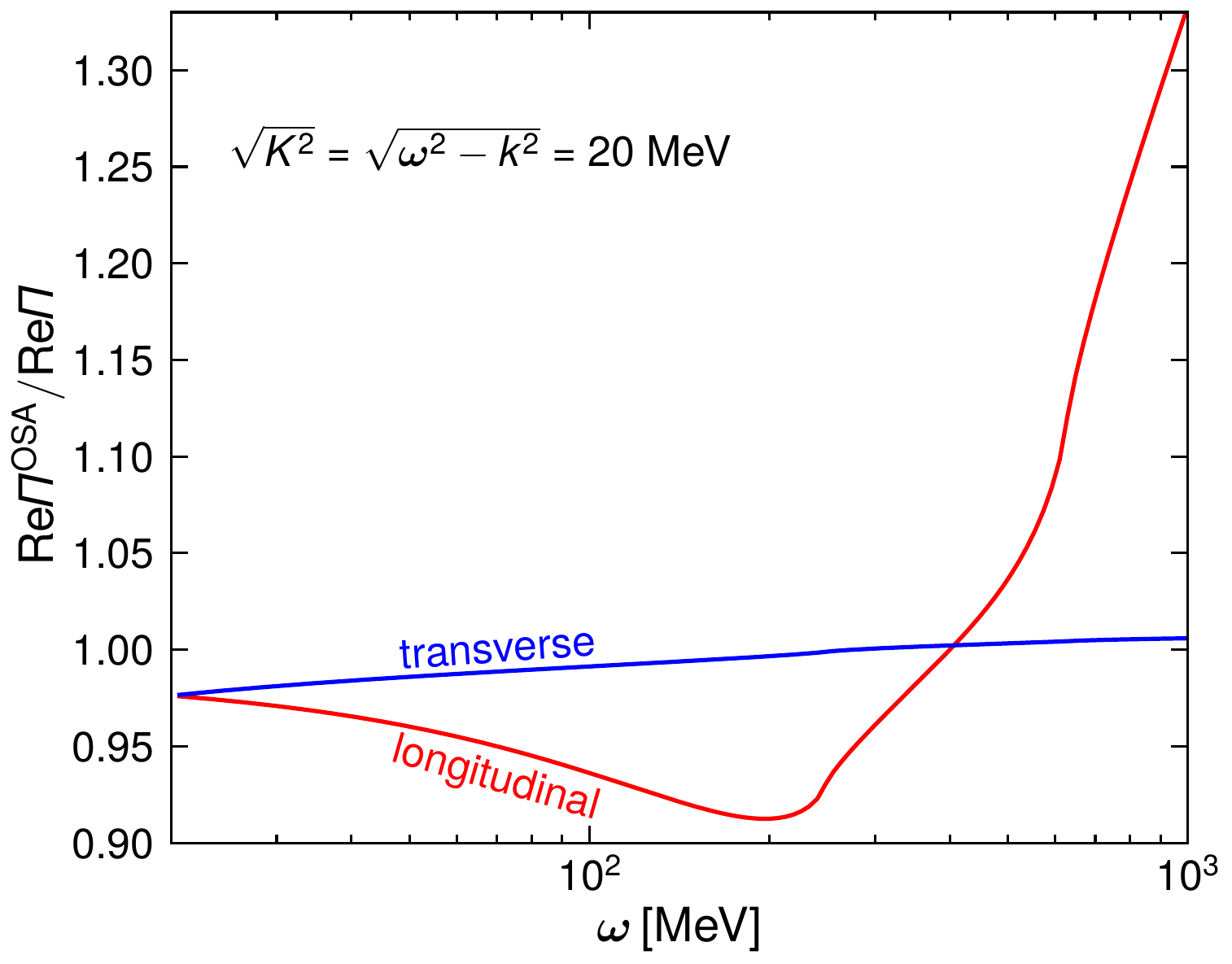} 
\caption{{\bf Left:} 
Real parts of $\Pi_{T}$ (blue) and $\Pi_{L}$ (red)
as a function of 
the photon energy $\omega$, 
both with full calculations (solid) \cite{Scherer:2024uui} 
and with the OSA (dashed) \cite{Raffelt:1996wa}. 
{\bf Right:} 
Ratio of Re$\Pi^{\rm OSA}$ to Re$\Pi$  
for the transverse (blue) and longitudinal (red) photons. 
On both figures, we have used the one-zone model and 
$\sqrt{K^2} = \sqrt{\omega^2 -k^2} = 20$ MeV.
}
\label{fig:s:RePi:compare3}
\end{figure}

In computing the real part of the 
EM polarization tensor $\Pi$, 
we use the results from Ref.~\cite{Braaten:1993jw}, where 
the $(K^2)^2/4$ term in the denominator of Eq.~\eqref{eq:pimunu} 
is dropped. 
Such an approximation is justified for on-shell photons,  
and we refer to it as the on-shell approximation (OSA). 
Moreover, the OSA is very useful, as it 
eliminates unphysical contributions 
from $\gamma \to e^+e^-$ 
for on-shell photons \cite{Braaten:1993jw}. 
However, the OSA should not be used without caution 
in the off-shell regime;
see Appendix E of Ref.~\cite{raffelt_stars_2023}
and
Ref.~\cite{Scherer:2024uui}. 
Recently, Ref.~\cite{Scherer:2024uui}  
computed the EM polarization tensor 
in the off-shell region and found that 
the OSA agrees with the full analysis in the relativistic limit. 
To explicitly check the agreement, 
in Fig.~\ref{fig:s:RePi:compare3}, 
we compare our calculations 
on the real part of $\Pi$ via the OSA 
with the full analysis in Ref.~\cite{Scherer:2024uui} 
in the off-shell region. 
The left-panel figure of 
Fig.~\ref{fig:s:RePi:compare3} shows that  
the real parts of $\Pi$ computed with the OSA
(denoted as ${\rm Re}\Pi^{\rm OSA}$) 
agree with the full analysis 
(denoted as ${\rm Re}\Pi$) 
from Ref.~\cite{Scherer:2024uui}, 
in the off-shell region where $\sqrt{K^2}=20$ MeV.  
The right-panel figure of Fig.~\ref{fig:s:RePi:compare3} shows that 
${\rm Re}\Pi^{\rm OSA}$ only deviates $\lesssim 3$\% 
from ${\rm Re}\Pi$ 
for the transverse mode when $\omega \lesssim 1$ GeV  
and $\lesssim 10$\% 
for the longitudinal mode when $\omega \lesssim 500$ MeV. 
Note that Re$\Pi_L$ in the $\omega \gtrsim 500$ MeV region 
is at least two orders of magnitude smaller than $K^2$ 
so that its contribution to the 
electron-positron annihilation 
cross section is $\lesssim 0.01$\%.
We thus conclude that the OSA is a good approximation 
in computing 
the real part of $\Pi$ 
in the off-shell region 
for the one-zone model.

\subsection{The imaginary part of the EM polarization tensor}
\label{appendix:imaginary}

The imaginary part of the EM polarization tensor is 
related to the photon absorption and production rates 
in the plasma \cite{Weldon:1982bn}. 
In the equilibrium case, 
one has ${\rm Im} \Pi=-\omega(1-e^{-\omega/T})\Gamma_{\rm abs}$ where $\Gamma_{\rm abs}$ is the photon absorption rate \cite{Weldon:1982bn}. 
In the SN core, 
the main contributions to 
the photon absorption rate that are relevant for 
the electron-positron annihilation process 
consist of the inverse-bremsstrahlung process 
of $\gamma p n \to p n$ \cite{Chang:2016ntp}, 
and the decay process of $\gamma \to e^+e^-$ \cite{Scherer:2024uui}.
\footnote{Because 
the photon is time-like with a positive energy
in the electron-positron annihilation process, 
the Cherenkov absorption process
where the photon is space-like
and the vacuum absorption process 
where the energy of the photon is negative \cite{Scherer:2024uui}  
are not relevant.} 
We compute the absorption rate 
due to the inverse-bremsstrahlung process 
for the transverse photon via \cite{Chang:2016ntp} 
\begin{align}
\Gamma_{\rm abs}^{T} = 
\frac{16\alpha n_n n_p}{3\pi \omega^3} 
\left(\frac{\pi T}{m_N}\right)^{3/2} \int_0^\infty x^2 
e^{-x} \sigma_{np}^T(x T) dx,
\end{align} 
and for the longitudinal photon via 
$\Gamma_{\rm abs}^{L}=(1-k^2/\omega^2)\Gamma_{\rm abs}^T$, 
where $m_N$ is the average nucleon mass, 
$n_n$ ($n_p$) is the neutron (proton) number density, 
$T$ is the temperature, 
$x\equiv E_{\rm cm}/T$ with $E_{\rm cm}$ being 
the non-relativistic center-of-mass energy of the $np$ system, 
and $\sigma_{np}^T$ is the transport cross section 
for the $n$-$p$ scattering. 
We use data from figure 3 of Ref.~\cite{Rrapaj:2015wgs} and 
figure 2 of Ref.~\cite{Brown:2018jhj} for $\sigma_{np}^T$. 
We compute the imaginary part of the EM polarization tensor due to 
$\gamma \leftrightarrow e^+e^-$ via \cite{Scherer:2024uui} 
\begin{equation}
    \text{Im} \Pi_a = -\frac{\alpha}{2 k}\int_{E_e^-}^{E_e^+} dE_e |\mathcal{M}_a|^2 (1 - f_{e^-} - f_{e^+}) 
    \Theta(K^2 - 4m_e^2)\Theta(\omega),
    \label{eq:ImPi:decay:1}
\end{equation}
where $E_e^{\pm} = (\omega \pm k \sqrt{1 - 4m_e^2/K^2})/2$, 
$|\mathcal{M}_L|^2 = K^2   \left( 4E_e\omega - 4E_e^2 - K^2 \right)/k^2$, and 
$|\mathcal{M}_T|^2 =  (K^2+2m_e^2)-|\mathcal{M}_L|^2/2$.

\begin{figure}[htbp]
\centering
\includegraphics[width=0.45 \textwidth]{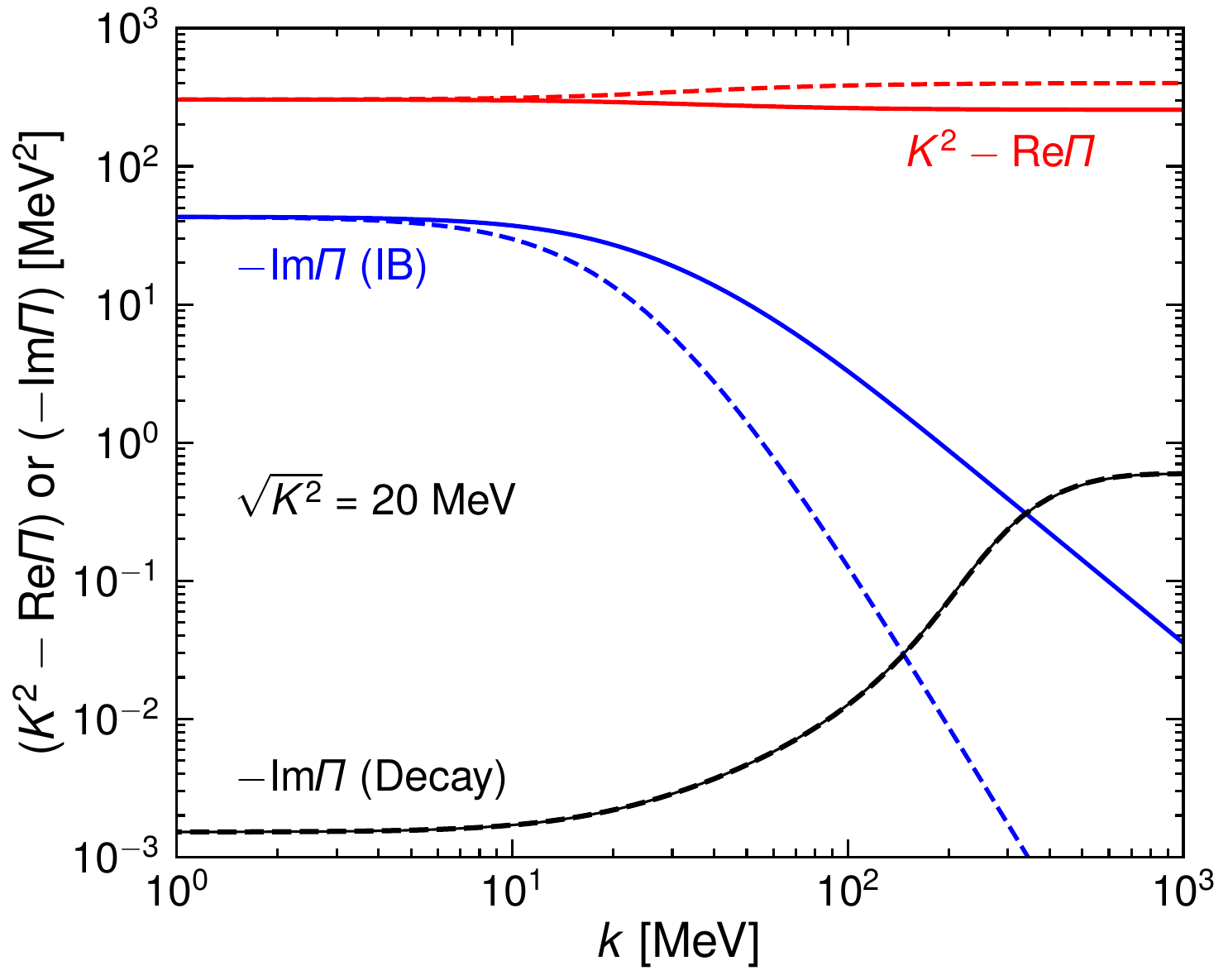} 
\hspace{0.5cm}
\includegraphics[width=0.45 \textwidth]{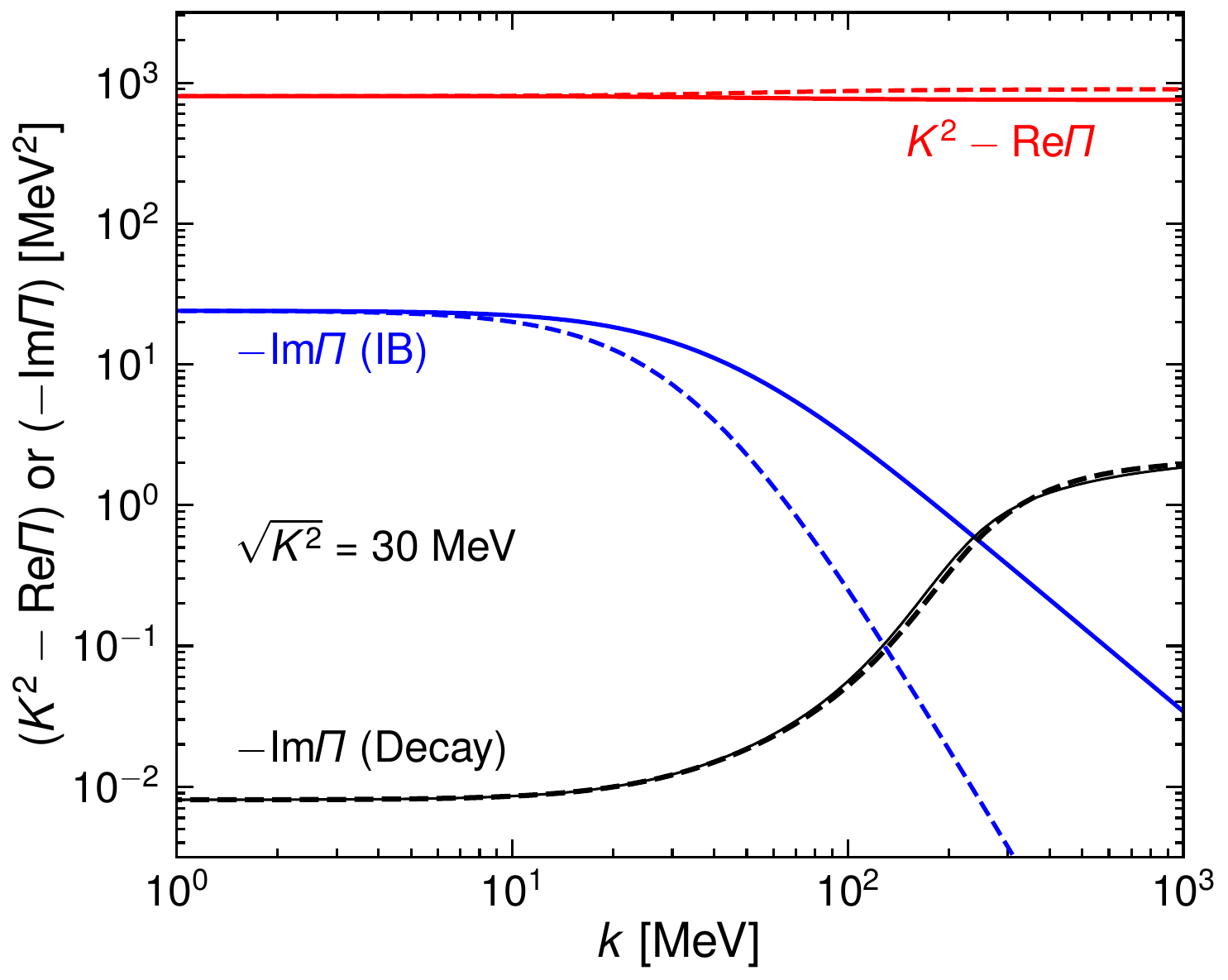}
\caption{{\bf Left:} 
The values of $(K^2-{\rm Re}\Pi)$ (red) \cite{Scherer:2024uui}, 
$-{\rm Im}\Pi$ due to the inverse bremsstrahlung process (blue) \cite{Chang:2016ntp}, 
and 
$-{\rm Im}\Pi$ due to the decay process (black) \cite{Scherer:2024uui}, 
as a function of the photon momentum $k$, 
both for the transverse mode (solid) 
and for the longitudinal mode (dashed). 
Here $\sqrt{K^2}=20$ MeV. 
{\bf Right:} The same as the left panel 
but with $\sqrt{K^2}=30$ MeV.} 
\label{fig:ImPi:compare}
\end{figure}

We compare the values of ($K^2-{\rm Re} \Pi$)
with $-{\rm Im} \Pi$ in 
the left-panel figure of
Fig.~\ref{fig:ImPi:compare} at $\sqrt{K^2}= 20$ MeV, 
which is near the threshold of $2m_e \simeq 18$ MeV. 
We find that the dominant contribution to ${\rm Im} \Pi$ 
comes from 
the inverse-bremsstrahlung process 
for $k\lesssim {\cal O}(100)$ MeV 
and 
the $\gamma \leftrightarrow e^+e^-$ process  
for $k\gtrsim 300$ MeV. 
In the low-$k$ (high-$k$) region, the magnitude of
${\rm Im} \Pi$ is at least 
$\sim 7$ (400) times smaller 
than ($K^2-{\rm Re} \Pi$), 
indicating a $\lesssim 2$\% contribution 
to the electron-positron annihilation cross section 
from the 
imaginary part of the EM polarization tensor.
We also find that 
as $\sqrt{K^2}$ moves further away from the 
threshold $2m_e \simeq 18$ MeV,
the
magnitude of
Im$\Pi$ 
due to the decay process 
increases,  
whereas 
the
magnitude of
Im$\Pi$ due to the bremsstrahlung process 
decreases, 
as shown in the 
right-panel figure of Fig.~\ref{fig:ImPi:compare}. 
As $(K^2-{\rm Re}\Pi)$ increases with increasing $K^2$, 
we find that the contributions from 
Im$\Pi$ at $\sqrt{K^2}=30$ MeV are even smaller 
than the $\sqrt{K^2}=20$ MeV case.

\section{Plasmon decay}
\label{appendix:plasmon:decay}

In this section we provide detailed calculations on 
the MCP flux from the plasmon decay process.

The spin-summed matrix element squared for the plasmon decay process,  
$\gamma \to \bar \chi \chi$, 
is given by
\begin{equation}
\sum_{\rm spins}|\mathcal{M}_a|^2=(\epsilon e)^2Z_{a}\epsilon_{a\mu}\epsilon_{a\nu}^*
T^{\mu\nu}(p_\chi, p_{\bar \chi}),
\label{eq:decay:rate:matrix}
\end{equation}
where
$e$ is the QED coupling, 
$\epsilon$ is the millicharge, 
and 
\begin{equation}
T^{\mu\nu} (p_\chi, p_{\bar \chi}) \equiv 4
\left( p^\mu_{\bar{\chi}} p^\nu_\chi + p^\mu_\chi p^\nu_{\bar{\chi}}-
p_\chi\cdot p_{\bar{\chi}}g^{\mu\nu}-m_\chi^2g^{\mu\nu} \right).
\end{equation} 
The plasmon decay rate 
in the plasma frame
is 
\begin{equation}
\Gamma_a =  \frac{1}{2\omega}   
(\epsilon e)^2Z_{a}\epsilon_{a\mu}\epsilon_{a\nu}^* 
\mathcal{I}^{\mu\nu}_\chi, 
 \label{eq:decay:rate}
\end{equation}
where 
\begin{equation}
    \mathcal{I}^{\mu\nu}_\chi
    \equiv\int d\Pi_\chi d\Pi_{\bar \chi}
    (2\pi)^4 
    \delta^{(4)}(K-p_\chi-p_{\bar{\chi}})T^{\mu\nu} 
    (p_\chi, p_{\bar \chi}), 
    \label{eq:decay:rate:I}
\end{equation}
with $d\Pi_\chi = d^3 p_\chi (2\pi)^{-3} (2E_\chi)^{-1}$. 
Carrying out the phase space integral, one finds \cite{Liang:2021kgw}
\begin{equation}
\mathcal{I}^{\mu\nu}_\chi
=\frac{K^2}{6\pi}f\left(\frac{m_\chi^2}{K^2}\right)P^{\mu\nu},
\label{eq:phasespaceintegral}
\end{equation}
where $P^{\mu\nu}\equiv-g^{\mu\nu}+{K^\mu K^\nu}/{K^2}
=P_T^{\mu\nu}+P_L^{\mu\nu}$ with 
$P_T^{\mu\nu} = \sum_{a=\pm}\epsilon_a^\mu\epsilon_a^{\nu*}$   
and $P_L^{\mu\nu} = \epsilon_L^\mu\epsilon_L^\nu$.
Due to the orthogonality of the polarization vectors, 
we obtain 
the plasmon decay width 
\begin{equation}
\Gamma_{a}   = 
Z_{a} \frac{\epsilon^2 \alpha K^2 }{3 \omega_{a}}
f\left( \frac{m_\chi^2}{K^2} \right), 
\end{equation}
for both the transverse $(a=T)$ 
and the longitudinal $(a=L)$ photons.

The number of photons 
decaying into the $\chi{\bar \chi}$ final state 
per unit volume per unit time is 
\begin{equation}
\frac{dN_a}{dVdt} = g_a\int\frac{d^3k}{(2\pi)^3}f_{\gamma}\Gamma_a,
\label{eq:plasmon:decay:rate:V}    
\end{equation}
where 
$g_T=2$, $g_L=1$, and 
$f_\gamma=1/(e^{\omega/T}-1)$ is the photon distribution.
In the photon rest frame, $\chi$ 
has an energy of $E_\chi^0 = \sqrt{K^2}/2$ 
and a polar angle $\theta$, 
where we have chosen the $z$ axis along the 
photon momentum in the plasma frame.
Thus, in the plasma frame, the energy of $\chi$ is   
\begin{equation}
E_\chi=\frac{\omega}{\sqrt{K^2}}\left(E_\chi^0+\frac{k}{\omega} 
\sqrt{(E_\chi^0)^2-m_\chi^2}
\cos\theta\right). 
\label{eq:chi:energy:plasma:frame}
\end{equation}
Since the distribution of $\chi$ is isotropic in the photon rest frame, 
in the plasma frame, $\chi$ is uniformly distributed in the energy  
range of $E_\chi^-\leq E_\chi\leq E_\chi^+$ with 
the maximal/minimal energy
$E_{\chi}^{\pm}= 
 \left( \omega \pm {k}
\sqrt{1-{4m_\chi^2}/{K^2}}
\right)/2$. 
Thus, the
production rate of MCPs per unit volume per unit energy 
in the plasma frame is given by
\begin{equation}
    \frac{d\Phi_a}{dE_\chi}=g_a\int\frac{d^3k}{(2\pi)^3}f_{\gamma}\Gamma_a
    g(E_\chi,m_\chi,K), 
\end{equation} 
where  
\begin{equation}
g(E_\chi,m_\chi,K) = 2\frac{\Theta(E_\chi-E_{\chi}^{-})\Theta(E_{\chi}^{+}-E_\chi)}{E_{\chi}^{+}-E_{\chi}^{-}}.
\end{equation}

\section{Proton bremsstrahlung}
\label{appendix:proton:bremsstrahlung}

In this section we provide detailed calculations on 
the MCP flux from the $np$ bremsstrahlung process.

The differential phase space for the 
$n(p_2)p(p_1)\to np\gamma(K)
\to n (p_4)p(p_3)\chi(p_\chi)\bar\chi(p_{\bar\chi})$ 
process 
can be decomposed as 
\begin{equation}
d\mathrm{LIPS}_4(p_{\mathrm{in}})=d\Pi_3d\Pi_4d\Pi_\chi d\Pi_{\bar\chi}(2\pi)^4\delta^{(4)}\left(p_\mathrm{in}-p_\mathrm{out}\right)= d\mathrm{LIPS}_3(p_{\mathrm{in}})\frac{dK^2}{2\pi}d\mathrm{LIPS}_2(K),
\label{eq:dLIPS}  
\end{equation}
where 
$p_\mathrm{in}=p_1+p_2$, 
$p_\mathrm{out}=p_3+p_4+p_\chi+p_{\bar\chi}$, and
$d\mathrm{LIPS}_3$ and $d\mathrm{LIPS}_2$ are the differential phase spaces for 
the $np\to np\gamma$ and the $\gamma\to \chi \bar \chi$ processes, respectively. 
The matrix amplitude for the 
$np\to np\chi\bar\chi$ process is given by
\begin{equation}
\mathcal{M}(np\rightarrow np\chi\bar\chi)=
\mathcal{M}_{\alpha}\frac{-ig^{\alpha\mu}}{K^2}j_{\mu},
\label{eq:MM:pb}
\end{equation}
where 
$j_{\mu}=i\epsilon e\bar{u}(p_{\chi})\gamma_{\mu}v(p_{\bar{\chi}})$, 
and 
$\mathcal{M}_{\alpha}$ is the matrix element 
for the $np$ part of the diagram. 
To avoid double counting with the plasmon decay process, 
we have neglected the plasma corrections to the photon propagator in 
Eq.~\eqref{eq:MM:pb} \cite{Chu:2019rok}. 
Then, the matrix amplitude squared,
averaged with initial spins and summed with final spins,
is given by
\begin{equation}
\frac{1}{4}
\sum_{\mathrm{spins}}|\mathcal{M}|^2 =
\frac{1}{4}
\sum_{\mathrm{spins}}\mathcal{M}^{\ast}_{\mu}\mathcal{M}_{\nu}\frac{\epsilon^2e^2}{(K^2)^2}T^{\mu\nu}(p_\chi,p_{\bar\chi}).\label{eq:matrixelement}
\end{equation}

Combining Eqs.~(\ref{eq:dLIPS}), (\ref{eq:matrixelement}), with (\ref{eq:phasespaceintegral}), 
we contract $\mathcal{M}_\mu^\ast\mathcal{M}_\nu$ with the photon polarization vector $\epsilon_a^\mu\epsilon_a^\nu$ 
to obtain the differential cross section of $np\rightarrow np\gamma$,
and thus the differential cross section of $np\rightarrow np\chi\bar\chi$ is \cite{Gninenko:2018ter,Liang:2021kgw,Du:2022hms}
\begin{align}
	\frac{d\sigma(np\rightarrow np\chi\bar\chi)}{dK^2d\omega}=\frac{\epsilon^2e^2}{12\pi^2}\frac{1}{K^2}\frac{d\sigma(np\rightarrow np\gamma)}{d\omega}f\left(\frac{m_\chi^2}{K^2}\right).
 \label{eq:brem:differentialxsec:app}
\end{align}

We then use the SRA to compute the cross section of $np\rightarrow np\gamma$:  
\begin{equation}
    \frac{d\sigma(np\rightarrow np\gamma)}{d\omega}= 
    \sigma^T_{np}
    \frac{d\mathcal{P}(K)}{d\omega},
    \label{eq:softphotonappro}
\end{equation}
where 
$\sigma^T_{np}$
is the transport cross section 
of the $(np\rightarrow np)$ process,
and $d\mathcal{P}(K)/d\omega$ is the photon splitting kernel.
In the SRA, $\mathcal{P}(K)$ is given by \cite{Chu:2019rok} 
\begin{equation}
    \mathcal{P}(K)=\frac{4\pi\alpha}{1-\cos\theta}
    \int\frac{d^3 k}{(2\pi)^32\omega}
    \sum_{\lambda}
    \left|J_p\cdot\epsilon_\lambda\right|^2, 
    \label{eq:splittingkernel}
\end{equation} 
where $\theta$ is the scattering angle of the proton in the center-of-mass frame, 
and 
\begin{equation}
J_p^\mu=
\frac{p_3^\mu}{p_3\cdot K}-\frac{p_1^\mu}{p_1\cdot K}. 
\end{equation}
Because 
$\sum_{\lambda}
\epsilon_\lambda^\mu 
\epsilon_\lambda^{\nu*}
= P^{\mu\nu}$ 
and $J_p \cdot K=0$, 
we have 
$\sum_{\lambda}
\left|J_p\cdot\epsilon_\lambda\right|^2
= - J_p^2$. 
In the non-relativistic limit, 
where $
m_p
\gg |{\bf p}_1|,|{\bf p}_3|$, 
we have
\begin{equation}
J_p^2 \simeq 
\frac{1}{{
m_p^2
\omega^2}}
\left[ 
\left( 
\frac{{\bf k}}{\omega} \cdot ({\bf p}_1-{\bf p}_3)
\right)^2
-({\bf p}_1-{\bf p}_3)^2
\right]. 
\end{equation}
After carrying out the integral over the solid angles of ${\bf k}$, 
we then obtain 
\begin{align}
\frac{d{\cal P}(K)}{d\omega} = 
\frac{4\alpha}{3\pi\omega}
\frac{E_{\rm cm}}{m_N}
f\left(\frac{K^2}{4\omega^2}\right). 
\end{align}

The production rate of MCPs per unit volume per unit energy 
in this process is given by 
\begin{align}
    \frac{d\Phi_{\rm pb}}{dE_\chi} = 
    g_1g_2\int \frac{d^3p_1}{(2\pi)^3} \frac{d^3p_2}{(2\pi)^3} f_1(E_1) f_2(E_2)
    \int d K^2 \int d\omega
    \frac{d\sigma (np\to np\chi\bar{\chi})}{dK^2d\omega} 
    |{\bf v}|_{\mbox{\scriptsize M\o l}}
    g(E_\chi,m_\chi,K), 
    \label{eq:Phibrem:1}
\end{align}
where 
$g_1=g_2=2$, 
$p_1$ ($p_2$) denotes the momentum of the initial proton (neutron), and
$f_1$ ($f_2$) is the Fermi-Dirac distribution of the initial proton (neutron).

Because in the SN core, $T\ll m_N$, we use 
Maxwell-Boltzmann distributions for the nucleons such that  
\begin{equation}
f_{1}(\textbf{p}_1)= 
\frac{n_{1}}{g_1}
 \left(\frac{2\pi}{
 m_1
 T}\right)^{3/2}
\exp \left( - \frac{{\bf p}_1^2}{2
m_1
T} \right),    
\end{equation}
and a similar expression for $f_2({\bf p}_2)$. 
It is convenient to use 
${\bf p}_s = ({\bf p}_1+{\bf p}_2)/\sqrt{2}$ 
and 
${\bf p}_r = ({\bf p}_1-{\bf p}_2)/\sqrt{2}$ 
such that $d^3p_1d^3p_2=d^3p_s d^3p_r$ and 
${\bf p}_1^2 + {\bf p}_2^2 = {\bf p}_s^2 + {\bf p}_r^2$. 
Because the integrand in 
Eq.~\eqref{eq:Phibrem:1} 
depends only on $|{\bf p}_r|$, we obtain \cite{Rrapaj:2015wgs} 
\begin{equation}
\int\frac{d^3p_1}{(2\pi)^3}\frac{d^3p_2}{(2\pi)^3}f_1(E_1)f_2(E_2)
=
\frac{n_1n_2}{g_1g_2}
\frac{2}{\sqrt{\pi T^3}}
\int dE_{\rm cm}
e^{-{E_{\rm cm}}/{T}}
\sqrt{E_{\rm cm}}, 
\label{eq:classical:integral}
\end{equation}
where 
$E_{\rm cm} = {\bf p}_r^2/2 m_N = m_N v_{\rm rel}^2/4$ 
with $m_N$ being the nucleon mass. This then leads to 
\begin{align}
\frac{d\Phi_{\rm pb}}{dE_\chi} =&
\frac{4 n_1n_2\epsilon^2\alpha}{3 \sqrt{m_N \pi^3 T^3}}
\int_{2m_{\chi}}^{\infty}dE_{\rm cm}E_{\rm cm}
e^{-{E_{\rm cm}}/{T}}\sigma^{{T}}_{np}(E_{\rm cm})
\nonumber\\
&\times\int_{4m_{\chi}^2}^{E_{\rm cm}^2} 
\frac{dK^2}{K^2} 
f\left(\frac{m_\chi^2}{K^2}\right)
\int_{\sqrt{K^2}}^{E_{\rm cm}} d\omega 
\frac{d\mathcal{P}(K)}{d\omega}g(E_\chi,m_\chi,K). 
\label{eq:productionrate:brem:1}
\end{align}
We note that 
Eq.~\eqref{eq:productionrate:brem:1} has 
an extra factor of 4 compared to 
equation 28 of Ref.~\cite{Chu:2019rok}.

\section{Electron-positron annihilation}
\label{appendix:electron:positron:annihilation}

In this section we provide detailed calculations on 
the MCP flux from the electron-positron annihilation process 
and the effective electron mass.

\subsection{MCP flux from the electron-positron annihilation process} 

The matrix element for 
the $e^+ e^- \to \gamma \to \bar \chi \chi$ process is 
\begin{equation}
    i\mathcal{M}= 
    \bar{u}(p_{\chi})(i\epsilon e\gamma_\mu)v(p_{\bar{\chi}}) 
    \tilde D^{\mu\nu} 
    \bar{v}(p_{2})(-ie\gamma_\nu)u(p_{1}), 
    \label{eq:ee:matrix}
\end{equation}
where $\tilde D^{\mu\nu}$ is the 
effective photon propagator given in Eq.~\eqref{eq:lorenz:propagator}.
The matrix element squared,
averaged with initial spins and summed with final spins, 
is given by 
\begin{equation}
\frac{1}{4}\sum_\mathrm{spins}|\mathcal{M_{\rm ann}}|^2=
\frac{\epsilon^2 e^4}{4} 
T^{\mu\rho} (p_1, p_2)
\tilde{D}_{\mu\nu}\tilde{D}_{\rho\sigma}^*  
T^{\nu\sigma}(p_\chi, p_{\bar\chi}).
\end{equation}
The cross section for 
the $e^+ e^- \to \gamma \to \bar \chi \chi$ process
is then given by 
\begin{equation}
\sigma_{\rm ann} =\frac{\pi^2  \epsilon^2\alpha^2}{E_1 E_2 |\textbf{v}|_{\mbox{\scriptsize M\o l}}} \tilde{D}_{\mu \nu} \tilde{D}^*_{\rho \sigma} T^{\mu \rho} (p_1, p_2) \mathcal{I}_\chi^{\nu \sigma},
\label{eq:anncross:section}    
\end{equation}
where $|\textbf{v}|_{\mbox{\scriptsize M\o l}} = 
\sqrt{K^2(K^2-4m_e^2)}/(2E_1E_2)$. 
Due to the orthogonality of the polarization vectors 
$\epsilon_a^\mu$, 
the annihilation cross section can be decomposed 
into the transverse and longitudinal components 
such that 
$\sigma_{\rm ann} = \sigma_{\rm ann}^T + \sigma^L_{\rm ann}$. 
The annihilation cross section with polarization $a$ is given by
\begin{equation}
\sigma^a_{\rm ann} = 
N_a
\frac{2\pi\epsilon^2\alpha^2}{3\sqrt{1-4m_e^2/K^2}}
\frac{K^2 }{(K^2-\mathrm{Re}\Pi_{a})^2+(\mathrm{Im}\Pi_a)^2}
    f\left(\frac{m_\chi^2}{K^2}\right),
    \label{eq:eeann:xsec2}
\end{equation}
where 
$N_L = 1-E_-^2/(E_+^2-K^2)$, 
$N_T = 1+4m_e^2/K^2+E_-^2/(E_+^2-K^2)$, 
where $E_\pm \equiv E_1 \pm E_2$ with 
$E_1$ ($E_2$) being the energy of the initial 
electron (positron).

The production rate of MCPs per unit volume per unit energy 
due to electron-positron annihilation is given by \cite{Chu:2019rok}
\begin{equation}
    \frac{d\Phi_{\rm ann}}{dE_\chi} = g_1 g_2
    \int 
   \frac{d^3p_{1}}{(2\pi)^3}\frac{d^3p_2}{(2\pi)^3} 
    f_{1}f_{2}
    \sigma_\mathrm{ann}
    |\textbf{v}|_{\mbox{\scriptsize M\o l}}g(E_\chi,m_\chi,K),
    \label{eq:phi:ee:B}
\end{equation} 
where $g_1=g_2=2$, 
and
$f_1$ ($f_2$) is the Fermi-Dirac distribution for 
the initial electron (positron). 
For an isotropic medium, one has 
\begin{equation}
	d^3p_1d^3p_2=8\pi^2 |\textbf{p}_1||\textbf{p}_2|E_1E_2dE_1dE_2d\cos\theta,\label{eq1:Jacobian}
\end{equation}
where $\theta$ is the angle between $\textbf{p}_1$ and $\textbf{p}_2$. 
It is convenient to use $E_\pm = E_1 \pm E_2$ and $K=p_1+p_2$ 
so that \cite{Chu:2019rok, Edsjo:1997bg}
\begin{equation}
	d^3p_1d^3p_2=2\pi^2E_1E_2dE_+dE_-dK^2.\label{eq:240717eq5.21}
\end{equation}
It is straightforward to find that 
$E_+\geqslant\sqrt{K^2}$ and $K^2\geqslant{\rm max}\{4m_e^2,4m_\chi^2\}$. 
Moreover, $|\cos\theta|\leqslant1$ leads to
\cite{Chu:2019rok, Edsjo:1997bg}
\begin{equation}
|E_-| 
\leqslant 
E_-^m \equiv 
\sqrt{1-\frac{4m_e^2}{K^2}}\sqrt{E_+^2-K^2}. 
\end{equation}
Substituting 
Eq.~\eqref{eq:240717eq5.21} 
into Eq.~\eqref{eq:phi:ee:B} leads to \cite{Chu:2019rok} 
\begin{equation}
    \frac{d\Phi_{\rm ann}}{dE_\chi} =
    \frac{1}{16\pi^4}
\int_{4m_{\mathrm{th}}^2}^{\infty}dK^2\int_{\sqrt{K^2}}^{\infty}dE_+\int_{-E_-^m}^{E_-^m}dE_- 
    f_{1}f_{2}\sqrt{K^2(K^2-4m_e^2)}\sigma_{\mathrm{ann}}g(E_\chi,m_\chi,K),
\label{eq:phi:ee:2:B}
\end{equation}
where $m_{\mathrm{th}}\equiv\max\{m_e,m_{\chi}\}$.

\subsection{Electron chemical potential}

The electron chemical potential, $\mu$, 
is related to the net electron density, $n_e$, 
which is given by  
$n_e = n_{e^-} - n_{e^+}$, where 
\begin{align}
n_{e^\pm}(T,\mu)
=&
\frac{1}{\pi^2}
\int_0^\infty
d|\textbf{p}|\,\textbf{p}^2
\frac{1}{e^{(E\pm\mu)/T}+1},
\label{eq:ChemicalPotential:electron}
\end{align}
where 
$E=\sqrt{\textbf{p}^2 + m_e^2}$ 
with $m_e$ being the electron mass, 
and 
$T$ is the temperature. 
For the one-zone model, 
where $T=30$ MeV and $n_e=2.72\times10^{-2}\ \mathrm{fm}^{-3}$
\cite{Caputo:2022mah}, 
the electron chemical potential is found to be $\mu \simeq 167$ MeV. 
\footnote{In the relativistic limit, one has 
$n_e \simeq \mu(\mu^2+\pi^2T^2)/(3\pi^2)$ \cite{Braaten:1993jw}, 
which is a good approximation for the one-zone model.}
One then finds $n_{e^+}/n_{e^-} \simeq 10^{-4}$, 
indicating a substantial positron number density. 
As a result, 
electron-positron annihilation can provide significant contributions 
to the production rate at high MCP masses.

\subsection{Effective electron mass}
\label{appendix:electron:mass}

Due to plasma effects, the 
electron mass in the SN core is modified so that 
it deviates significantly from its vacuum value, 
$m_e^0 = 0.511$ MeV. 
Thus, in the electron-positron annihilation cross section, 
Eq.~\eqref{eq:eeann:xsec2}, 
one has to use the effective electron mass to properly account 
for the plasma effects.

The effective mass for an electron that has a four-momentum $p^{\mu}=(E,\textbf{p})$ 
in the medium is given by 
\begin{equation}
	m_e^{\mathrm{eff}} (\textbf{p})=\sqrt{(m_e^0)^2-2(A_{\gamma}+A_e)-2m_e^0(C_{\gamma}+C_e)}, 
 \label{eq:effectivemass}
\end{equation}
where $A_{\gamma}$, $A_e$, $C_{\gamma}$, and $C_e$ 
are the four functions that describe the mass corrections due to plasma effects \cite{Hardy:1998if,Chu:2019rok}. 
In a neutral medium where the positron number density is not negligible, 
the four functions in Eq.~(\ref{eq:effectivemass}) are given by \cite{Hardy:1998if}
\begin{align}
	A_e=&\frac{-\alpha}{4\pi|\textbf{p}|}\int_{0}^{\infty}dq\frac{|\textbf{q}|}{\sqrt{\textbf{q}^2+(m_e^0)^2}}\nonumber\\
    &
    \times\left\{4|\textbf{p}||\textbf{q}|\left[f_1(\textbf{q})+f_2(\textbf{q})\right]
    +\left[(m_e^{\mathrm{eff}})^2+(m_e^0)^2\right]\left[f_2(\textbf{q})L_1-f_1(\textbf{q})L_2\right]\right\},\label{eq:eq5.43}\\
	C_e=&\frac{\alpha m_e^0}{\pi|\textbf{p}|}\int_{0}^{\infty}dq\frac{|\textbf{q}|}{\sqrt{\textbf{q}^2+(m_e^0)^2}}\left[f_2(\textbf{q})L_1-f_1(\textbf{q})L_2\right],\\
	A_{\gamma}=&\frac{-\alpha}{4\pi|\textbf{p}|}\int_{0}^{\infty}dqf_{\gamma}(\textbf{q})\left\{8|\textbf{p}||\textbf{q}|+\left[(m_e^{\mathrm{eff}})^2+(m_e^0)^2\right](L_3-L_4)\right\},\\
	C_{\gamma}=&\frac{\alpha m_e^0}{\pi|\textbf{p}|}\int_{0}^{\infty}dqf_{\gamma}(\textbf{q})(L_3-L_4),
\end{align}
where $\textbf{q}$ is the three-momentum of medium particles (electron, positron, photon) 
and $L_{1,2,3,4}$ are functions of $\textbf{q}$ \cite{Chu:2019rok,Hardy:1998if}:
\begin{align}
	&L_1(\textbf{q})=\ln\left[\frac{2\left(E\sqrt{\textbf{q}^2+(m_e^0)^2}+|\textbf{p}||\textbf{q}|\right)-(m_e^{\mathrm{eff}})^2-(m_e^0)^2}{2\left(E\sqrt{\textbf{q}^2+(m_e^0)^2}-|\textbf{p}||\textbf{q}|\right)-(m_e^{\mathrm{eff}})^2-(m_e^0)^2}\right],\\
	&L_2(\textbf{q})=\ln\left[\frac{2\left(E\sqrt{\textbf{q}^2+(m_e^0)^2}+|\textbf{p}||\textbf{q}|\right)+(m_e^{\mathrm{eff}})^2+(m_e^0)^2}{2\left(E\sqrt{\textbf{q}^2+(m_e^0)^2}-|\textbf{p}||\textbf{q}|\right)+(m_e^{\mathrm{eff}})^2+(m_e^0)^2}\right],\\
	&L_3(\textbf{q})=\ln\left[\frac{2|\textbf{q}|(E+|\textbf{p}|)+(m_e^{\mathrm{eff}})^2-(m_e^0)^2}{2|\textbf{q}|(E-|\textbf{p}|)+(m_e^{\mathrm{eff}})^2-(m_e^0)^2}\right],\\
	&L_4(\textbf{q})=\ln\left[\frac{2|\textbf{q}|(E+|\textbf{p}|)-(m_e^{\mathrm{eff}})^2+(m_e^0)^2}{2|\textbf{q}|(E-|\textbf{p}|)-(m_e^{\mathrm{eff}})^2+(m_e^0)^2}\right],
\end{align}
where $f_1(\textbf{p})$, $f_2(\textbf{p})$, and $f_{\gamma}(\textbf{p})$ are thermal momentum distributions of electrons, positrons, and photons, respectively. 
Note that the right-hand side of Eq.~(\ref{eq:effectivemass}) also contains $m_e^{\mathrm{eff}}$. 
By solving Eq.~(\ref{eq:effectivemass}), 
we find that the effective electron mass is $m_e^{\mathrm{eff}}(\textbf{p})\approx9\,\mathrm{MeV}$ 
in the one-zone model. 
We then use $m_e^{\mathrm{eff}}(\textbf{p})\approx9\,\mathrm{MeV}$ 
for the electron mass in computing the 
electron-positron annihilation cross section.

\section{SN cooling limit}
\label{appendix:SN:cooling:limit}

In this section 
we describe our analysis of the SN cooling limits  
by taking into account the electron-positron annihilation process, 
which was often neglected due to the scarcity of positrons 
in the SN; see e.g., Ref.~\cite{Chang:2018rso}. 
However, we find that 
the electron-positron annihilation process 
is the dominant production process 
for MCPs with mass $\gtrsim$ 57 MeV.   
Thus, it is of great importance to compute the 
SN cooling limits by taking into account 
the electron-positron annihilation process. 
To do so, we compute the MCP luminosity (energy emitted per unit time) 
by taking into account both the proton bremsstrahlung 
and the electron-positron annihilation processes:  
\begin{equation}
    L_\chi=4\pi\int_0^{R_\nu}dr r^2(\dot{Q}_{\rm pb}+\dot{Q}_{\rm ann}),
\end{equation}
where $L_\chi$ is the MCP luminosity, $R_\nu$ 
is the radius of the neutrinosphere, 
and $\dot{Q}_{\rm pb}$ ($\dot{Q}_{\rm ann}$) is the energy loss rate of the proton bremsstrahlung (electron-positron annihilation) process, which is given by
\begin{equation}
    \dot{Q}_{\rm pb/ann}=\int_{m_\chi'}^\infty dE_\chi\frac{d\Phi_{\rm pb/ann}}{dE_\chi}E_\chi.
\end{equation}
Note that the plasmon decay process is unimportant 
in the high-mass regime. 
We obtain the limits by 
requiring $L_\chi\leq L_\nu=3\times10^{52}$ erg/s. 
In order to compare with the SN cooling limits 
in Ref.~\cite{Chang:2018rso}, 
we adopt the fiducial model given in Ref.~\cite{Chang:2018rso} 
for the SN model.

\section{LESN constraints on MCPs with large coupling}
\label{appendix:large:coupling}

In this section we describe our procedure to obtain the 
upper boundary of the LESN exclusion region on MCPs, 
which occurs at sufficiently large $\epsilon$.

For sufficiently large $\epsilon$ values, 
MCPs form a blackbody sphere (referred to as the MCP-sphere) 
within the SN; 
see e.g., Ref.~\cite{Chang:2018rso}. 
In this case, MCPs are emitted from the surface of the MCP-sphere
with a blackbody spectrum: 
\be
\frac{d\Psi_\chi}{dE_\chi} = \frac{g_\chi}{8\pi^2} \frac{E_\chi^2 - m_\chi^2}{e^{E_\chi/T(R_\chi)} + 1},
\ee
where $R_\chi$ is the radius of the MCP-sphere
and $g_\chi=4$ accounts for the spin degree of freedom of both $\chi$ and $\bar\chi$. 
To determine $R_\chi$, we adopt the analysis of the neutrinosphere in 
Ref.~\cite{Cooperstein:1988fz}. 
Thus, we use \cite{Cooperstein:1988fz} 
\be
t_{\rm diff}(R_\chi) = t_{\rm dyn}(R_\chi),
\ee
where 
\begin{equation}
t_{\rm diff}(R_\chi) = \frac{3R_\chi^2}{\pi^2 \lambda_\chi} 
\end{equation}
is the diffusion time for randomly walking MCPs in a uniform sphere, 
and 
\begin{equation}
t_{\rm dyn}(R_\chi) = \left( 0.1 \sqrt{6\pi G \rho(R_\chi)} \right)^{-1}
\end{equation}
is the dynamical time associated with the collapse to a singular point.  
Here $\lambda_\chi$ is the mean free path of MCPs \cite{Davidson:2000hf}, 
$ \rho$ is the mass density, 
and $G$ is the gravitational constant. 
The total energy deposition outside the neutrinosphere is given by 
\be
E_m = 4\pi R_\chi^2 \Delta t\int dE_\chi \frac{d\Psi_\chi}{dE_\chi} \Delta E_\chi, 
\label{ErgDep:blackbody}
\ee
where $\Delta t = 3$ s, and 
$\Delta E_\chi$ is the energy deposited by a single $\chi$ particle outside the neutrinosphere.

\section{Pionic process}
\label{appendix:PionicProcess}

In this section,
we provide detailed calculations
on the MCP flux
from the process of 
$\pi^-p\rightarrow n\chi\bar\chi$.

Because the only SM particle that $\chi$ 
couples to is the photon, 
the cross section of 
$\pi^-p\rightarrow n\chi\bar\chi$ 
is related to the 
process of $\pi^-p\rightarrow n \gamma$. 
There are three diagrams for $\pi^-p\rightarrow n \gamma$: 
these can be obtained by removing the final states 
$\chi$ and $\bar\chi$ particles 
in the three diagrams in Fig.~\ref{fig:PionProcess}.
The amplitude of $\pi^-p\rightarrow n\gamma$ 
for these three diagrams are 
\begin{align}
&\mathcal{M}_1=-i\frac{eg_A}{\sqrt{2}f_\pi}\epsilon_\mu^\ast(k)\frac{1}{2p_p\cdot k}\bar u(p_n)(2p_p^\mu\slashed{p}_\pi-\slashed{p}_\pi\slashed{k}\gamma^\mu)\gamma^5 u(p_p),
\label{eq:MatrixElement:1}
\\
&\mathcal{M}_2
=
i\frac{eg_A}{\sqrt{2}f_\pi}\epsilon_\mu^\ast(k)
\bar u(p_n)\gamma^\mu\gamma^5u(p_p),
\\
&\mathcal{M}_3
=i\frac{eg_A}{\sqrt{2}f_\pi}\epsilon_\mu^\ast(k)\bar{u}(p_n)\gamma^5 u(p_p)\frac{2m_Np_\pi^\mu}{p_\pi\cdot k+(m_\pi^2-p_\pi^2)/2}.
\label{eq:MatrixElement:3}
\end{align}
Note that 
Eqs.~(\ref{eq:MatrixElement:1})--(\ref{eq:MatrixElement:3}) 
agree with Eqs.~(26)--(28) in Ref.~\cite{Shin:2022ulh} 
except that the former contain 
an extra factor of $g_A/2$. 
The total matrix element is 
$\mathcal{M}_\mathrm{tot}=\mathcal{M}_1+\mathcal{M}_2+\mathcal{M}_3$. 
Thus one has 
\begin{align}
&\left\langle
\sum_{\mathrm{spins},\lambda}
|\mathcal{M}_\mathrm{tot}|^2
\right\rangle
=
\frac{8e^2g_A^2m_N^2}{f_\pi^2}
\left\{
1+\frac{p_\pi^2}{\tilde{\omega}^2}
\left[
\frac{\tilde{\omega}^2}{2\omega^2}
+
\frac{m_\pi^2\tilde{\omega}^2}{2(\tilde{\omega}^4-\textbf{p}_\pi^2\omega^2)}
-
\frac{\mathrm{arctanh}(|\textbf{p}_\pi|\omega/\tilde{\omega}^2)}{|\textbf{p}_\pi|\omega/\tilde{\omega}^2}
\right]
\right\},
\label{eq:AngularAverage:SquaredMatrixElement}
\end{align}
where we have 
summed the initial and final nucleon spins, 
as well as final photon polarizations, 
and averaged over the solid angle $d\Omega=2\pi d\cos\theta$
where $\theta$ is the scattering angle of 
the outgoing photon with respect to the 
incoming pion. 
Note that we have expanded 
the results 
up to the order of $\mathcal{O}(m_N^2)$ in
Eq.~\eqref{eq:AngularAverage:SquaredMatrixElement}. 
Note that 
Eq.~\eqref{eq:AngularAverage:SquaredMatrixElement} 
differs from Eq.~(32) in Ref.~\cite{Shin:2022ulh},
not only by an additional factor of $g_A^2/4$,
but also the term $\tilde{\omega}^2/(2\omega^2)+m_\pi^2\tilde{\omega}^2/[2(\tilde{\omega}^4-\textbf{p}_\pi^2\omega^2)]$ in the square brackets, 
which is one in Eq.~(32) of Ref.~\cite{Shin:2022ulh}.

The production rate of MCPs
per unit volume per unit energy
due to the $\pi^-p\rightarrow n\chi\bar\chi$ process is
\begin{align}
\frac{d\Phi_\mathrm{pion}}{dE_\chi}
= &
\int
d\Pi_\pi
d\Pi_p
d\Pi_n
d\Pi_\chi
d\Pi_{\bar\chi}
f_\pi f_p(1-f_n)
g(E_\chi,m_\chi,K)
\sum_\mathrm{spins}|\mathcal{M}_\chi|^2
\nonumber\\
&\times(2\pi)^4\delta^{(4)}(p_p+p_\pi-p_n-p_\chi-p_{\bar\chi}), 
\label{eq:EnergyLossRate:chi}
\end{align}
where $\mathcal{M}_\chi
=\mathcal{M}_\gamma^\mu(-ig_{\mu\nu}/K^2) j^\nu$ 
with $\mathcal{M}_\gamma^\mu$ being defined via 
$\mathcal{M}_\mathrm{tot}\equiv 
\mathcal{M}_\gamma^\mu \epsilon_\mu^\ast$, 
and $j^\nu=i\epsilon e\bar{u}(p_{\chi})\gamma^\nu v(p_{\bar{\chi}})$.

Following the calculations in Appendix \ref{appendix:proton:bremsstrahlung},
we decompose 
the differential phase space
for
$\pi^-(p_\pi)p(p_p)\rightarrow n(p_n)\gamma(K)\rightarrow n(p_n)\chi(p_\chi)\bar\chi(p_{\bar\chi})$
as follows 
\begin{equation}
d\mathrm{LIPS}_5(p_\mathrm{in})
=
d\Pi_\pi d\Pi_p d\Pi_n d\Pi_\chi d\Pi_{\bar\chi}
(2\pi)^4\delta^{(4)}(p_\mathrm{in}-p_\mathrm{out})
=
d\mathrm{LIPS}_4(p_\mathrm{in})
\frac{dK^2}{2\pi}d\mathrm{LIPS}_2(K),
\end{equation}
where
$p_\mathrm{in}=p_\pi+p_p$,
$p_\mathrm{out}=p_n+p_\chi+p_{\bar\chi}$,
$d\mathrm{LIPS}_5$ 
and $d\mathrm{LIPS}_4$
are the differential phase spaces
for the
$\pi^-p\rightarrow n\chi\bar\chi$
and
$\pi^-p\rightarrow n\gamma$
processes, respectively.

Thus, we have 
\begin{equation}
\int d\mathrm{LIPS}_5(p_\mathrm{in})
\sum_\mathrm{spins}|\mathcal{M}_\chi|^2
=
\frac{\epsilon^2e^2}{24\pi^2}
\int d\mathrm{LIPS}_4(p_\mathrm{in})
\int_{4m_\chi^2}^{E_\pi^2}
\frac{dK^2}{K^2}
f\left(
\frac{m_\chi^2}{K^2}
\right)
\left\langle
\sum_{\mathrm{spins},\lambda}
|\mathcal{M}_\mathrm{tot}|^2
\right\rangle. 
\end{equation}
Then 
Eq.~(\ref{eq:EnergyLossRate:chi}) becomes
\begin{align}
\frac{d\Phi_\mathrm{pion}}{dE_\chi}
=&\frac{\epsilon^2e^4g_A^2}{96}
\sqrt{\frac{2(m_N^\ast)^3T_c^9}{\pi^{14}f_\pi^4}}z_pz_\pi
\int_0^\infty dx_p\frac{x_p^2}{e^{x_p^2}+z_p}\frac{e^{x_p^2}}{e^{x_p^2}+z_n}
\int_0^\infty dx_\pi\frac{x_\pi^2}{e^{\kappa_\pi-y_\pi}}
g(E_\chi,m_\chi,x_\pi)
\nonumber\\
&\times
\int_{4m_\chi^2}^{E_\pi^2}\frac{dK^2}{K^2}
f\left(
\frac{m_\chi^2}{K^2}
\right)
\sqrt{1-\frac{K^2}{E_\pi^2}}\textbf{m}^2. 
\end{align}

\section{Magnetic field effect}
\label{appendix:magnetic:field}

In this section we discuss the effects of the magnetic 
fields in the SN core on the constraints on MCPs. 
Due to magnetic flux conservation 
during the evolution from a proto-neutron star to 
a neutron star, the magnetic fields 
in the SN core can become significant, 
reaching strengths of $10^{10}$ G \cite{Powell:2022ljv}, or even higher. 
Such large magnetic fields 
could potentially prevent the MCPs from freely streaming. 
This can be easily seen via 
the gyroradius: 
\begin{equation}
    r_g =\frac{\gamma m_\chi v_\chi}{\epsilon e B},
\end{equation}
where 
$B$ is the magnetic field, 
$v_\chi$ is the MCP velocity, 
and $\gamma=1/\sqrt{1-v_\chi^2}$ is the Lorentz factor. 
We find $r_g \simeq $ 2.2 m 
for 
$(m_\chi, E_\chi, \epsilon) 
= (100\;\mathrm{MeV}, 125\;\mathrm{MeV}, 10^{-7})$.

However, the gyroradius only describes 
the motion in the plane 
that is perpendicular to the magnetic field. 
In SN explosions 
MCPs are emitted 
in all directions. 
To study the magnetic effects on MCPs, 
we thus carry out a Monte Carlo simulation where 
the MCP momentum can point to any direction.

For the magnetic field of the SN core,  
we adopt the dipole field profile
\cite{Suwa:2007nq,Obergaulinger_2014,Matsumoto:2020rbz}: 
\be
A_r = A_\theta = 0,\qquad
A_\phi = \frac{B_0}{2}\,\frac{r_0^{3}}{r^{3}+r_0^{3}}\,r\sin\theta,
\ee
where $A_{r,\theta,\phi}$ is the vector potential in the $r,\theta,\phi$ directions, respectively, $r$ is the radius. 
For the initial magnetic field configuration, 
we adopt 
$r_0=1000$ km, which is the radius of the core, 
and $B_0$ in the range of 
$(10^{8}-10^{12})$ G \cite{Obergaulinger_2014}.
Then at 1 second post bounce, the magnetic field should 
take the following form: 
\begin{equation}
A_r = A_\theta = 0,\qquad
A_\phi = \frac{B_{\rm{PNS}}}{2}\,\frac{r_{\rm{PNS}}^{3}}{r^{3}+r_{\rm{PNS}}^{3}}\,r\sin\theta, 
\end{equation}
where $r_{\rm{PNS}}=12.9$ km is the radius of the PNS core,  
and 
\begin{equation}
    B_{\rm PNS} 
    \sim 10^{15}\, {\rm G}
    \left(
    \frac{B_0}{10^{12}\, {\rm G}}
    \right)
    \left(
    \frac{30\, {\rm km}}{r_{\rm PNS}}
    \right)^2,
\end{equation}
which is due to the magnetic flux conservation.

In our MC analysis, 
we simulate 
$10^{4}$ MCPs that are emitted from the PNS center,  
where the motion of MCPs 
is governed by the Lorentz-force law: 
\begin{equation}
\frac{d\textbf{p}_\chi}{dt} = \epsilon e\, \textbf{v}_\chi \times \textbf{B}, 
\label{eq:lorentz_force}   
\end{equation}
where $\textbf{p}_\chi = \gamma m \textbf{v}_\chi$ 
is the MCP momentum, and 
$\textbf{v}_\chi$ is the MCP velocity. 
For each particle, 
we compute the point in the phase space 
for every ns 
and follow the trajectory for 1 ms.

In our analysis, we consider the following five  
different $B_0$ values: 
$10^8$, 
$10^9$, 
$10^{10}$, 
$10^{11}$, 
and $10^{12}$ G. 
We find that for the benchmark model point:  
$(m_\chi, E_\chi, \epsilon)$ = 
$(100\; {\rm MeV}, 125\; {\rm MeV}, \\ 10^{-7})$, 
more than 99\% of the simulated MCPs 
escape the neutrino sphere, 
which is at $r=40$ km, 
for the five different values of $B_0$.

Thus, although the gyroradius is small compared to the 
PNS radius, a significant fraction of MCPs are able to 
escape the neutrinosphere. 
This is in part due to the high energy of MCPs, 
so that 
any initial momentum component parallel to the magnetic field 
quickly carries MCPs outward, allowing them to escape. 
We thus conclude that the strong magnetic fields 
in the SN core 
do not significantly hinder the free streaming of MCPs.

\bibliography{ref.bib}{}

\bibliographystyle{utphys28mod}

\end{document}